\documentclass[aps,prl,twocolumn,superscriptaddress]{revtex4-1}
\usepackage{graphics}
\usepackage{picture}
\usepackage{xcolor}
\usepackage{color}
\usepackage{tikz}

\begin{document}

%\title{Controlling the Light Emission Frequency of a Two-Level System %with Optical Pulses}
\title{Suppressing Spectral Diffusion of the Emitted Photons with Optical Pulses}

\author{H. F. Fotso}
\affiliation{Department of Physics and Astronomy, Iowa State University, Ames, Iowa 50011, USA}
\author{A. E. Feiguin}
\affiliation{Department of Physics, Northeastern University, Boston, Massachusetts 02115, USA}
\author{D. D. Awschalom}
\affiliation{Institute for Molecular Engineering, University of Chicago, Chicago, IL 60637, USA}
\author{V. V. Dobrovitski}
\email{slava@ameslab.gov}
\affiliation{Ames Laboratory US DOE, Ames, Iowa, 50011, USA}

\begin{abstract}
In many quantum architectures the solid-state qubits, such as quantum dots or color centers, are interfaced via emitted photons. However, the frequency of photons emitted by solid-state systems exhibits slow uncontrollable fluctuations over time (spectral diffusion), creating a serious problem for implementation of the photon-mediated protocols. Here we show that a sequence of optical pulses applied to the solid-state emitter can stabilize the emission line at the desired frequency. We demonstrate efficiency, robustness, and feasibility of the method analytically and numerically. Taking nitrogen-vacancy (NV) center in diamond as an example, we show that only several pulses, with the width of 1 ns, separated by few ns (which is not difficult to achieve) can suppress spectral diffusion. Our method provides a simple and robust way to greatly improve the efficiency of photon-mediated entanglement and/or coupling to photonic cavities for solid-state qubits.
\end{abstract}


\maketitle

%==========BODY OF PAPER =========================================

The ability to transfer quantum information between the stationary qubits via photons is at the heart of many applications such as long-range quantum networks and quantum interface between distant qubits \cite{KimbleQuInternet,DLCZ,ChildressRepeater,BernienHanson,
PfaffHanson,QDTeleport}. The photon-mediated entanglement is based on %interference of 
indistinguishable photons (having the same polarization and frequency) emitted by two different stationary qubits and entangled with them \cite{ChildressRepeater,BarrettKok,BernienHanson,PfaffHanson}. 
It is of central importance for such solid-state qubits as quantum dots and color centers, which are often difficult to couple directly, while the photon-mediated protocols present a very promising alternative \cite{BernienHanson,PfaffHanson,QDTeleport}. At low temperatures, a noticeable fraction of photons emitted from these qubits is concentrated in the zero-phonon line (ZPL) and is insensitive to the phonon absorption/emission. The photons emitted into the ZPL are naturally entangled to the originating solid-state qubits \cite{ToganLukin,BuckleyAwsch,NVzpls2,QDTeleport,CarterGammonQDcavity,SantoriVuckovicYamamoto,QDSpinPhotEnt}, and constitute excellent flying qubits; 
the emission into the ZPL can be enhanced by placing the qubit into a cavity \cite{HuetalSiVinNanocav,FaraonEtalNatPhot}.
%,FaraonEtalPRL}. 

However, ensuring indistinguishability of the photons emitted by two different quantum dots or color centers remains a crucial challenge \cite{FuEtal,BassettAwschElTunNV,
%BernienHansonPRL,
QDspdif1,QDspdif2,QDspdif3,BernienHanson,PfaffHanson,HansomAtature,AcostaBeausoleil}. Changes in the local strain and motion of the charges around the emitter lead to slow random variation (spectral diffusion) of the energies of the levels involved in the photon emission. The position of the ZPL (i.e.\ the frequency of the emitted photons) fluctuates with the amplitude far exceeding the natural linewidth. Thus, the spectral overlap between the photons coming from two different qubits is greatly reduced, resulting in low efficiency of the heralded entanglement process. The same problem occurs when the qubit is coupled to the photonic cavity: due to spectral diffusion of the ZPL, the overlap of the emitted photons with the cavity line is diminished, thereby reducing the Purcell enhancement. Due to severity of the problem, solutions have been actively sought, and the schemes based e.g.\ on active feedback \cite{BassettAwschElTunNV,HansomAtature,AcostaBeausoleil,PrechtelEtal}, 
three-level emitters coupled to the cavities \cite{SantoriBeausoleil,HeEtalPRL}, special emission regimes \cite{MattAtatureNCommResFl,MattAtaturePRLResFl}, have been explored.

Here we suggest a conceptually simple, general, and robust protocol for suppressing the spectral diffusion of the ZPL of the solid-state emitters. 
Since the frequency of the emitted light is determined by the average phase accumulated between the states of the emitter over the spontaneous emission time, one can modify the emission spectrum by changing the phase between the relevant states with optical pulses. 
Below we show 
that by applying a series of short optical control pulses to the solid-state emitter, the center of the zero-phonon emission line can be pinned at any desired frequency, determined by the carrier frequency of the pulses; this is demonstrated both analytically and numerically. Taking NV centers in diamond as an example, we show that only several pulses of 1~ns width (corresponding to the optical Rabi frequency of 0.5~GHz), separated by 5--6~ns, are sufficient to suppress the spectral diffusion of the ZPL. The protocol is robust with respect to small non-idealities of the pulses, and is explicitly shown to significantly improve the photon indistinguishability. Our work shows how the emission spectrum can be engineered using the pulse protocol, despite fast internal dynamics of the photon bath. Further exploring this venue can be of much interest for a wide class of problems concerning photon emission.

\begin{figure}
%\begin{center}
%\includegraphics*[width=8.0cm]{schematic.pdf}
%\includegraphics[angle=270,width=8cm]{Fig1v4.eps}
\includegraphics[angle=270,width=\columnwidth]{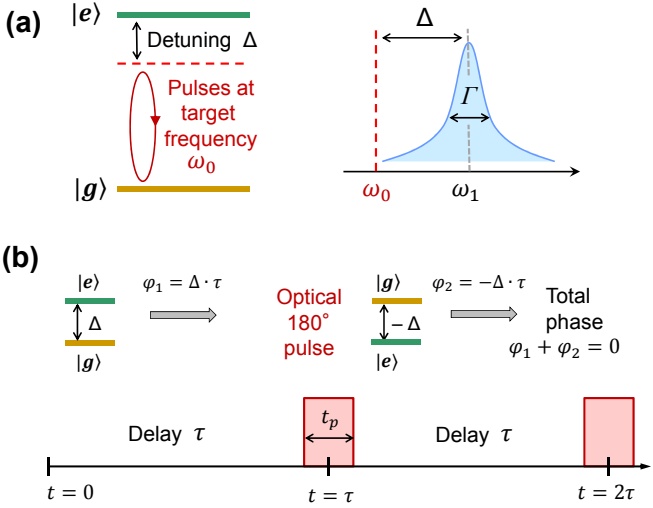}
\caption{(Color online) (a) Excited state of the two-level system (solid-state emitter) is shifted by random amount $\Delta$ from the desired position $\omega_0$, so that the spontaneous emission line (of width $\Gamma$) is centered at $\omega_1=\omega_0+\Delta$. To shift the line to the target frequency, a sequence of pulses with the carrier frequency $\omega_0$ is applied.
(b) The optical 180$^\circ$ control pulses are applied periodically, with the interval $\tau$. In the rotating frame, each pulse swaps the ground and the excited state, reversing the detuning $\Delta\to -\Delta$. The total phase accumulated before and after the pulse is nullified, and emission happens as if the detuning was absent, with ZPL centered at $\omega_0$.} 
%\label{fig:problem_schematic}
\label{fig:schem}
%\end{center}
\end{figure}

We model the solid-state emitter as a two-level system, emitting photons in the course of spontaneous transition from the excited state $|e\rangle$ (where the emitter initially is), located at the energy $\hbar\omega_1$ above the ground state $|g\rangle$ (below we set $\hbar=1$), see Fig.~\ref{fig:schem}(a). The optical control pulses, each of very short duration $t_p$, are applied at the carrier frequency $\omega_0$, so it is convenient to work in the rotating-wave approximation (RWA), using the basis rotating with the frequency $\omega_0$. The system's Hamiltonian then has the form 
\begin{equation}
\label{eq:hamiltonian}
H = H_c(t) + \frac{\Delta}{2} \sigma_z - i\sum_{k=0}^{L-1} g_{k} \left( a^{\dagger}_{k} \sigma^- - a_{k} \sigma^+ \right) + \sum_{k=0}^{L-1} \omega_k a^{\dagger}_{k} a_k,
\end{equation}
where $\Delta=\omega_1-\omega_0$ is the detuning of the ZPL from the target frequency, caused by the random fluctuation in the local strain or charge environment; this detuning is static on the spontaneous emission timescale. The operators $\sigma_z=|e\rangle\langle e|-|g\rangle\langle g|$, $\sigma^+ =|e\rangle\langle g|$, and 
$\sigma^- =(\sigma^+)^\dagger$ describe the emitter, 
$a_k$ is the annihilation operator of the $k$-th photon mode ($L$ modes in total), $g_k$ is its coupling strength, and $\omega_k$ is its detuning from $\omega_0$. The Hamiltonian $H_c(t)$, describing the control pulses, can be taken as $H_c(t)=(\Omega/2)[\sigma^+ + \sigma^-]$ during the pulses and zero otherwise; for ideal (instantaneous, 180$^\circ$) pulses $\Omega=\pi/t_p$ and $t_p\to 0$
(experimentally, the optical Rabi frequency $\Omega$ should be large in comparison with the typical $\Delta$). 
During the pulses, under strong driving $\Omega\ll\Gamma$, the incoherent scattering is dominant \cite{Milonni}.  
%We do not consider the coherently scattered light during the pulses, %its intensity : in experiments this light is disregarded or screened %away, and only the fluorescence photons are to be detected.
%
By including the RWA directly in the Hamiltonian, we assume that $\omega_1$ is appropriately renormalized \cite{AckerhaltEberly,RF_KimbleMandel77}, and the non-Markovian effects \cite{Milonni,MarkovApprox_WodEberl} in the electromagnetic bath can be neglected.

Our approach is based on a qualitative argument that the frequency of the emitted light is determined by the {\em average} rate of phase accumulation \cite{AckerhaltEberly,RF_KimbleMandel77,OptPulsesKhitrin} between the states $|e\rangle$ and $|g\rangle$ over the time of spontaneous emission $t_0$. 
This is due to the fact that on the timescales short in comparison with the time $t_0$ the emitter and the emitted radiation constitute a single coherently evolving quantum system, and the properties of the emitted photon are determined by the whole history of what happened to the emitter during the spontaneous emission time, not only by its instant condition. In our case, by applying the optical control pulses, the average (over the timescale $t_0$) rate of the phase accumulation is modified, because 
each pulse changes $\sigma_z$ to $-\sigma_z$, so the detuning term $(\Delta/2)\sigma_z$ changes its sign, see Fig.~\ref{fig:schem}. If several pulses are applied within the time $t_0=1/\Gamma$ then the average detuning is nullified, and the appropriately averaged accumulated phase corresponds to the emission frequency $\omega_0$. Below, we assume a simple periodic pulse pattern with the inter-pulse delay $\tau$, as shown in Fig~\ref{fig:schem}(c). 

There is a similarity between our approach and the dynamical decoupling (DD) method, which has been used to decouple various quantum systems from their environment \cite{DD1,DD2,DD3,DD4}. 
However, in contrast with the standard optical pulse DD \cite{Kurizki,Walther}, the control pulses here do not attempt to cancel the coupling of the qubit to the electromagnetic bath; this would require extremely short \cite{Walther} inter-pulse delay $\tau\lesssim\omega_0^{-1}$ and would suppress emission altogether. Instead, we  use the pulses to cancel the detuning, and in this way  redirect emission from some electromagnetic modes to others. It may also be possible to achieve the same effect with the continuous control of the emitter, in analogy to the continuous-wave decoupling \cite{cwd1,cwd2,cwd3,cwd4}, and consider the sequences with other pulse timings \cite{SupMat}: this could provide novel ways of modifying the properties of the emitted photons, and constitute an interesting topic for future research.

We characterize the emission spectrum via the number of photons of a given frequency $\omega$:
$N_{\omega}(t) = \sum'_k \langle a_k^{\dagger}(t) a_k(t) \rangle$, where summation is over the modes with $\omega_k=\omega$; note explicit dependence on time $t$. 
Within RWA description, the relevant frequencies are confined to the vicinity of $\omega_0$, so that $\omega_k\in [-D,D]$ where $D\ll\omega_{0,1}$ but much larger than all other frequency scales of the problem. Within this region the coupling parameters for all modes are practically constant, $g_k=g$ for all 
$k=0,\dots , L-1$, and the photon density of states $\rho_\omega$ is also constant. Thus we choose $\omega_k=-D+k\epsilon$, with $\epsilon=2D/(L-1)$; with this choice $\rho_\omega=1/\epsilon$. 
In reality $L\to\infty$, which implies the scaling $g\propto L^{-1/2}$ and $\epsilon\propto L^{-1}$. 
We also assume fixed polarization of the emitted photons \cite{FuEtal,PfaffHanson}. Without control pulses, the solution is the standard Lorentzian emission line \cite{RF_Heitler,AckerhaltEberly} centered at $\Delta$ with the width at half maximum $\Gamma=2\pi g^2 \rho_\omega$. Everywhere below we normalize energy and time by $\Gamma$ and $t_0=\Gamma^{-1}$, respectively, setting $\Gamma=2$.

We consider the problem using both analytical and numerical approaches in parallel.
For analytical treatment we employ the standard approach based on the weak coupling/Markov approximation, used for studying spontaneous decay and resonant fluorescence \cite{AckerhaltEberly,RF_KimbleMandel77,RF_Heitler,
MarkovApprox_WodEberl}. 
We use the toggling Heisenberg representation: between the pulses the operators $\sigma_z(t)$, $\sigma^{\pm}(t)$, and $a_k(t)$ evolve according to standard Heisenberg representation, while the control pulses change the emitter operators $\sigma^\pm\to\sigma^\mp$,  $\sigma^z\to -\sigma^z$. The corresponding equations of motion for the time-dependent operators after the $n$-th pulse are
\begin{eqnarray}
\label{eq:eom_1}
\dot a_k &=& -i \omega_k a_k + g_k \left( \xi_1 \sigma^- + \xi_2 \sigma^{\dagger} \right) \\ \nonumber
%\label{eq:eom_2}
\dot \sigma^- &=&  -i (-1)^n \Delta \sigma^- + \sum_k g_k \xi_1 a_k \sigma_z - \sum_k g_k \xi_2 a^{\dagger}_k \sigma_z\\ \nonumber 
%\label{eq:eom_3}
\dot \sigma_z &=& -2\sum_k g_k \left[ \xi_1 a^{\dagger}_k \sigma^- - \xi_2 a^{\dagger}_k \sigma^+ - \xi_2 a_k \sigma^- + \xi_1 a_k \sigma^+ \right],
\end{eqnarray}
where we introduced the periodic functions $\xi_1(t)$ and $\xi_2(t)$ of period $2\tau$, such that $\xi_1(t)=1$ for $t<\tau$ (before the pulse) and $\xi_1(t)=0$ for $\tau<t<2\tau$ (after the pulse), 
while  $\xi_2(t)=1-\xi_1(t)$. The equations of motion (\ref{eq:eom_1}) can be integrated iteratively between consecutive pulses using the Markovian approximation \cite{AckerhaltEberly,MarkovApprox_WodEberl}, and the answer can be obtained in the limit of the large number of pulses; see Supplemental Material \cite{SupMat} for details.

\begin{figure}
\includegraphics[width=8cm]{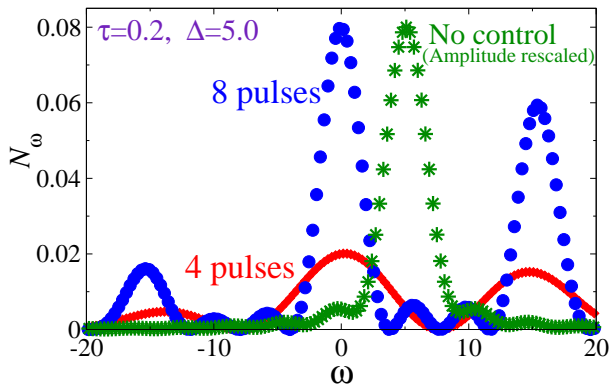}
\caption{(Color online) Emission profile $N_\omega$ in the presence of control pulses, for $\tau=0.2$ and $\Delta=5.0$, after 4 pulses (red diamonds) and 8 pulses (blue circles), as compared with the free spontaneous emission spectrum (green stars). Amplitude of the latter is rescaled for easier comparison with the 8-pulse spectrum.} 
\label{fig:problem_solution}
\end{figure}

The analytically calculated fluorescence spectra $N_{\omega}(t)$ are shown in Fig.~\ref{fig:problem_solution} for $\Delta=5.0$. The free emission (no control pulses) spectrum is compared with the pulse-controlled emission for the inter-pulse delay $\tau=0.2$. The total number of emitted photons increases with the number of pulses, so the amplitude of the no-control spectrum has been rescaled. The spectra agree with our qualitative arguments. The no-control ZPL is centered at $\omega=\Delta$, and only a tiny fraction of emission is present at the target frequency $\omega=0$. In contrast, the control pulses shift the ZPL to the target position. Although additional satellite peaks appear on the sides, about 50\% of the spectral weight is successfully moved to $\omega=0$ peak. The emission peaks are wide at $t<t_0$, and acquire their natural width $\Gamma$ at longer times.

\begin{figure}
\includegraphics[width=8cm]{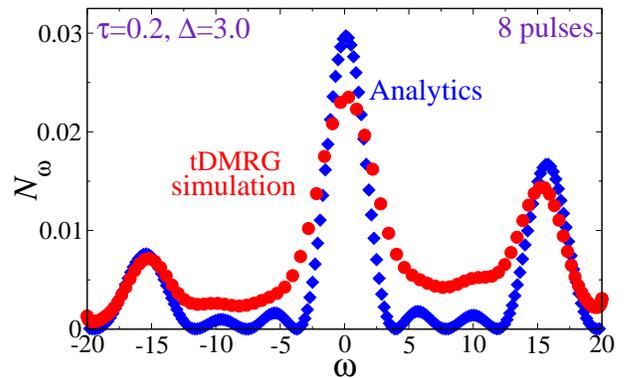}
\caption{(Color online) Emission profiles $N_\omega$ in the presence of control pulses, for $\tau=0.2$ and $\Delta=3.0$, after the 8-th pulse. 
Blue diamonds represent the analytical results, while the red circles represent the spectrum obtained via time-dependent density matrix renormalization group (tDMRG) simulations. The latter is rescaled by a factor $2/\pi$ in order to take into account different density of the photon modes between the analytical model and the 1-D chain used in tDMRG simulations.} 
\label{fig:simulation_theory}
\end{figure}

We used numerical simulations to independently check analytical  approximation, and to investigate the impact of the pulse imperfections. Each pulse increases the number of total excitations in the system, $n_{ex}=(1+\sigma_z)/2 + \sum_k a^\dagger_k a_k$, leading to an exponential increase in the number of relevant states with time. To make the problem tractable, we model the photonic bath in a different way, as a periodic 1-D chain of $L$ harmonic oscillators, with the site 0 coupled to the two-level system (emitter); the corresponding Hamiltonian is
\begin{eqnarray}
\label{eq:hamiltonian2}
H &=& H_c(t) + \frac{\Delta}{2} \sigma_z + C \left( \sigma^+ b_0 + \sigma^-b_0^{\dagger} \right) \nonumber \\
&& -i(D/2)\sum_{j=0}^{L-1} \left( b_jb_{j+1}^{\dagger} -b_j^{\dagger}b_{j+1}  \right)
\end{eqnarray}
where $b_j^{\dagger}$ and $b_j$ are the creation and annihilation operators for a boson at site $j$, respectively. Using Fourier transform of the bosonic operators, it is easy to see that this Hamiltonian is equivalent to Eq.~\ref{eq:hamiltonian}, provided that $g_k=g=C/\sqrt{L}$ and $\omega_k=D\sin{2\pi k/L}$. The latter dispersion relation ensures that the density of states in the vicinity of the emission line (near $\omega_k=0$) is also constant, $\rho_\omega=[\pi D/L]^{-1}$ (note the double degeneracy of each $\omega_k$), and the value of $g$ is adjusted to ensure $\Gamma=2$. The increased density of states at the edges (near $\omega_k=\pm D$) is irrelevant because $N_\omega(t)$ is small there.
Using this model for the photonic bath, the problem can be efficiently solved for large values of $L$ and long times (large number of control pulses), using the time-dependent density matrix renormalization group (tDMRG) method \cite{tDMRG} using symmetries to reduce the entanglements introduced by the periodic boundary conditions \cite{tDMRG2}.

The two methods, analytics vs.\ tDMRG, are compared in Fig.~\ref{fig:simulation_theory}. Good agreement between the two approaches is clearly seen, taking into account the different photon dispersion laws and couplings $g$, the used analytical approximations (weak coupling, large number of pulses), and despite the fact that the parameters ($L=201$, $D=20$, $\rho_\omega\approx 3.2$) are far from the ideal quasi-continuous broad spectrum of the photons with $L\gg 1$, $D\gg 1$, and $\rho_\omega\gg 1$. In order to account for different spectral density of the photon modes ($\rho_\omega=L/(2D)$ for analytics and $\rho_\omega=L/(\pi D)$ for tDMRG), the tDMRG simulation results for $N_\omega$ are multiplied by a factor $2/\pi$.
%(e.g.\ in the simulations only 18 photon frequencies $\omega_k$ fall %within the ZPL region $\omega\in[-\Gamma/2,\Gamma/2]$). 

%
\begin{figure}
\includegraphics*[width=8cm]{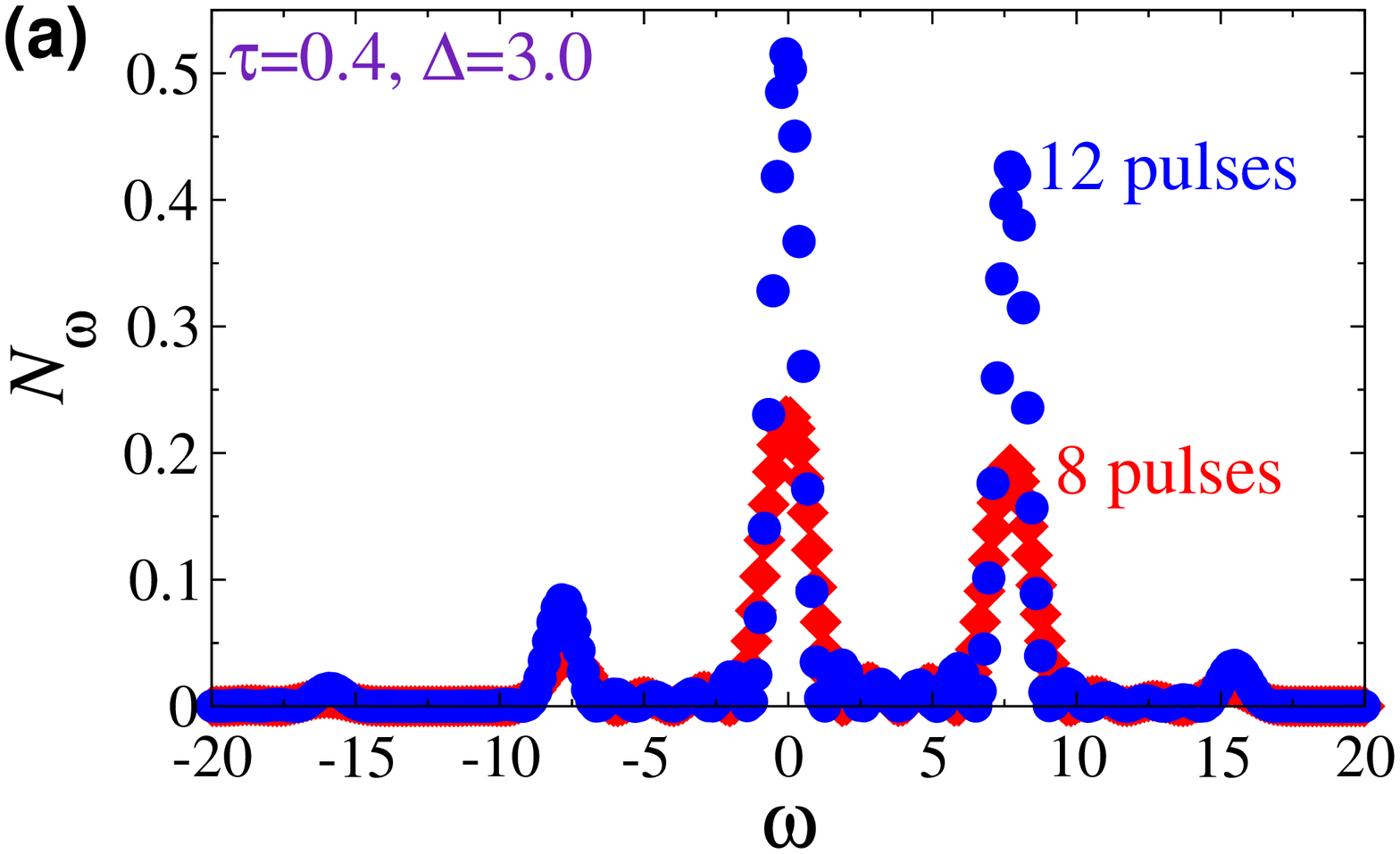} 
\includegraphics*[width=8cm]{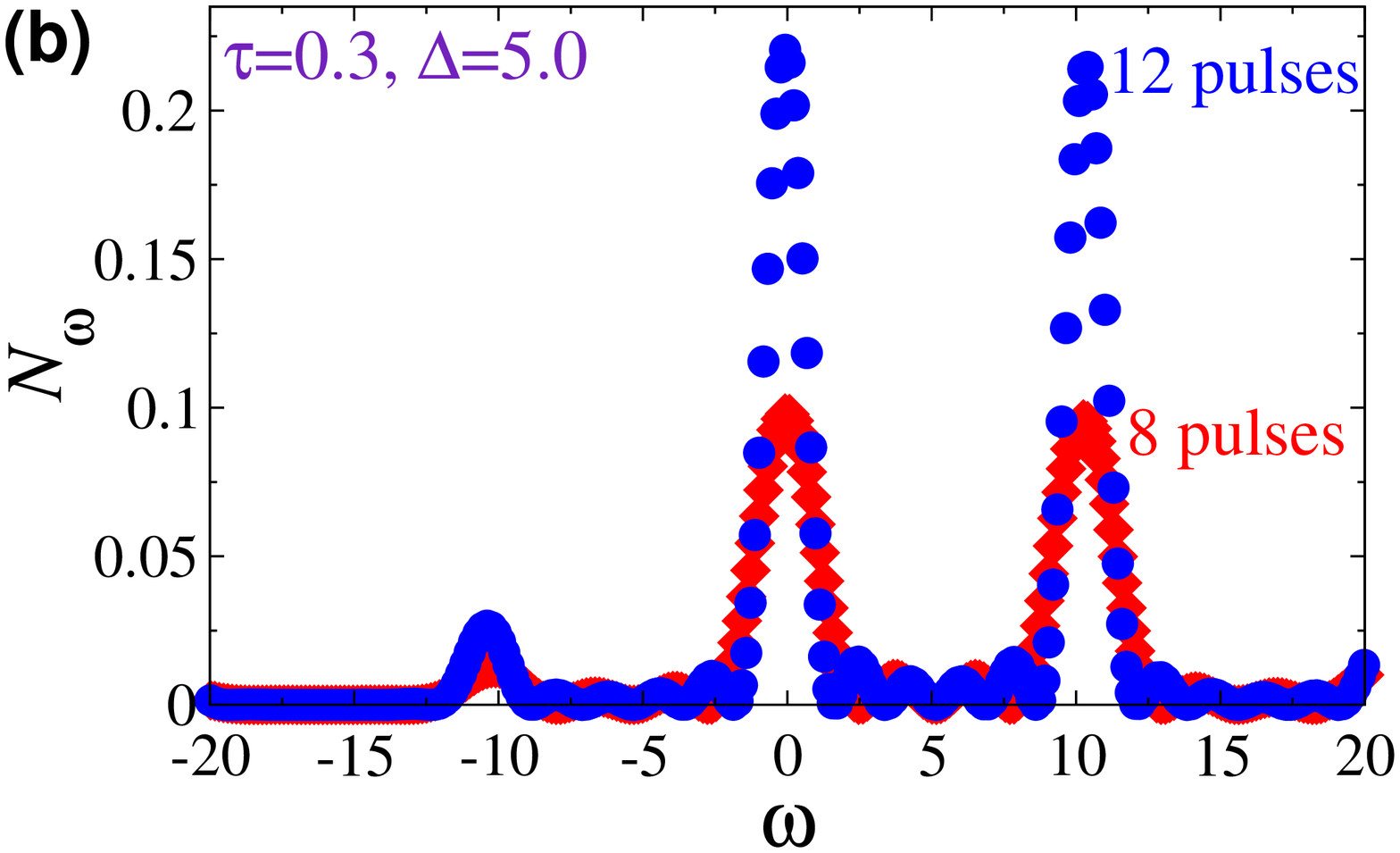} 
\caption{(Color online) 
Emission profiles $N_\omega$ after 8 pulses (red diamonds) and 12 pulses (blue circles). (a)  $\tau=0.4$ and $\Delta=3.0$,  The satellite peaks at $\omega=\pm \pi/\tau$ and $\pm 2\pi/\tau$ are clearly seen. (b) $\tau=0.4$ and $\Delta=5.0$. Both graphs show that the protocol works for large delays $\tau>1/\Delta$, delivering about 50\% of emission to the target frequency.} 
\label{fig:satelite_evolution}
\end{figure}

Dependence of the controlled emission profile on $\Delta$ and $\tau$ is shown in Fig.~\ref{fig:satelite_evolution}. As expected, the central peak at $\omega=0$ is flanked with the satellite peaks at $\omega=\pm \pi/\tau, \pm 3\pi/\tau,\dots$, since each pulse produces a 180$^\circ$ phase rotation. While the emission into these satellites is unwanted, a large fraction of the spectral power is still retained in the central peak. It is important that our protocol does not require very short inter-pulse delays $\tau$, and works even when $\tau >\Delta^{-1}$, so that even large detunings can be eliminated with moderately spaced pulses. The overall structure of the emission profile remains unchanged even at larger $\tau$. Only, the spectral weight of the central peak decreases for $\tau > 2\Delta^{-1}$, while the satellite peak with the frequency closest to $\Delta$ grows \cite{SupMat}.

%
%\begin{figure}[htbp]
%%\includegraphics*[width=8.0cm]
%{simultnImperfFiniteWidthDelta3p0tau0p2_Np6.pdf}  
%\includegraphics[width=8cm]
%{simultnImperfFiniteWidthDelta3p0tau0p2_Np6_4.eps}  
%\caption{(Color online) Emission spectra $N_\omega$ for $\tau=0.2$ %and $\Delta=3.0$ after 6 pulses. The results for ideal, instant %180$^\circ$, pulses (blue circles) are compared with those for %imperfect pulses: red diamonds --- instant pulses with incomplete %rotations, 175$^\circ$ instead of ideal 180$^\circ$; green triangles %--- pulses of finite width $t_p=0.05$. All three spectra practically %coincide, showing robustness of the protocol.} 
%\label{fig:nonIdealPulses}
%\end{figure}
%%

Finally, we tested robustness of the approach with respect to two most typical experimental non-idealities, the incomplete rotation during the optical control pulses, and the finite width of the control pulses. 
%
%We compared the results of the tDMRG simulations for non-ideal and %for ideal (instantaneous 180$^\circ$) pulses. 
We find that a moderate 5$^\circ$ error in the rotation angle does not affect efficiency of the control. In the same way, pulses as wide as $t_p=0.05$ (which is 1/4 of the inter-pulse distance $\tau$) remain as efficient as ideal pulses. The corresponding spectra
%(practically coinciding with the ideal-pulse spectra) 
are given in the Supplemental Material. 
Thus, the requirement of sufficiently large optical Rabi driving, $\Delta\ll\Omega=\pi/t_p$, which is needed to ensure that the rotation is close to 180$^\circ$, would not be difficult to satisfy.

By suppressing the spectral diffusion, our protocol improves indistinguishability of the emitted photons. We analyzed the coincidence count rate for the two-photon interference experiments, and the results show significant improvement: informally speaking, about half of the photons become indistinguishable when the pulse control is applied to the emitter; the detailed calculations are presented in the Supplemental Material.

As a specific example, let us consider a nitrogen-vacancy (NV) center in diamond, which has several ZPL separated by 3--5~GHz, corresponding to different excited orbital levels \cite{FuEtal,NVzpls,NVzpls2,RogersNJP,PfaffHanson}.
At low temperatures \cite{FuEtal,ZPL_LowT} the ZPL has the natural width $\Gamma=2\pi\cdot 16$~MHz, corresponding to the spontaneous decay time $t_0=10$~ns. The typical range of the detuning fluctuations $\Delta\sim 5 \Gamma = 2\pi\cdot 100$~MHz, so that only a small portion of emission occurs at the target frequency $\omega=0$. However, if the control pulses of duration $t_p=0.05$ are applied, separated by $\tau=0.3$, then about 50\% of the emission goes into the central line at $\omega=0$. For NV centers, this corresponds to the inter-pulse delay $\tau=6$~ns and the pulse width $t_p=1$~ns, i.e.\ optical Rabi driving $\Omega=\pi/t_p=2\pi\cdot 0.5$~GHz. These parameters are easily achievable in comparison with the typically used optical Rabi driving of few GHz and sub-ns timing. The modest Rabi driving also limits ionization of NV center, and ensures that other ZPLs, located several GHz away, are not affected. 

Concluding, we suggested and analyzed a pulse protocol for suppression of spectral diffusion of the zero-phonon line of a solid-state emitter, which is one of the central problems on the way to implementing the long-range quantum networks with solid-state nodes. We demonstrated feasibility and robustness of the protocol. This approach is simple, does not involve additional levels, and avoids long delays associated with feedback methods (but can also be used along with the latter for fine tuning of the ZPL). 
More generally, our results show that the pulse control can be efficiently used to manipulate even fast (Markovian) environments, where the typical intra-bath evolution times are far shorter than the inter-pulse delays and pulse durations. Exploring this venue of quantum control can develop solutions for a wide class of problems concerning bosonic and fermionic environments. 

We thank L. C. Bassett, R. Hanson, and T. H. Taminiau for inspiring discussions. This work was partially supported by AFOSR MURI program and NSF. The work at Ames Laboratory (design and analysis of the protocol) was supported by the  Department of Energy --- Basic Energy Sciences under Contract No. DE-AC02-07CH11358. A.E.F.\ acknowledges NSF support through grant DMR-1339564.

%\end{document} 

\end{document}

\title{Supplemental Material for:\\
Suppressing Spectral Diffusion of The Emitted Photons with Optical Pulses
}

\author{H. F. Fotso}
\affiliation{Department of Physics and Astronomy, Iowa State University, Ames, Iowa 50011, USA}
\author{A. E. Feiguin}
\affiliation{Department of Physics, Northeastern University, Boston, Massachusetts 02115, USA}
\author{D. D. Awschalom}
\affiliation{Institute for Molecular Engineering, University of Chicago, Chicago, IL 60637, USA}
\author{V. V. Dobrovitski}
\affiliation{Ames Laboratory US DOE, Ames, Iowa, 50011, USA}

\maketitle

\section{Model and Pulse Sequence}

We consider a two-level atom (spin-$\frac{1}{2}$ particle) which is coupled to a bath of photons.
This problem is described by the following Hamiltonian \cite{Milonni}:
\begin{equation}
H=\sum_{\lambda,\bf k} \hbar\omega_{\bf k} a^{\dagger}_{\lambda,\bf k}a_{\lambda,\bf k} + (\hbar\omega_1/2) \sigma_z -i \sum_{k} g_{\lambda,\bf k} \left( \sigma^- + \sigma^+ \right) \left( a_{\lambda,\bf k} - a^{\dagger}_{\lambda,\bf k}\right)
\end{equation}
where $\hbar\omega_1$ is the distance between the two levels involved in the ZPL photon emission, the operators $\sigma_z$ and $\sigma^{\pm}$ describe the two-level emitter, and $a^\dagger_{\lambda,\bf k}$ is the creation operator for the photon mode with polarization $\lambda$ and momentum $\bf k$, so that $\omega_{\bf k}=|{\bf k}|c$. The coupling constant between the emitter and  the photon of the mode $(\lambda, \bf k)$ is 
\begin{equation}
g_{\lambda,\bf k} = \omega_1 ({\bf d}{\bf e}_\lambda) \sqrt{\frac{2\pi\hbar}{\omega_{\bf k} V}}
\end{equation}
where $\bf d$ is the dipolar matrix element of the transition, ${\bf e}_\lambda$ is the polarization of the photon mode, and $V$ is the normalization volume. Since we consider only emission with the fixed polarization, we can drop the index $\lambda$ below; we also set $\hbar=1$ and $c=1$ for convenience of notation.

We employ RWA description, considering only the relevant photon modes in the vicinity of the target frequency $\omega_0$, so that the frequencies $\omega_{\bf k}$ are confined to the interval $[\omega_0-D,\omega_0+D]$ where $D\ll\omega_0$. At the same time, $D$ is much larger than all other relevant frequency scales of the problem, i.e.\ $D\gg \Delta,\ \Gamma,\ \pi/\tau$ (here $\Delta=\omega_1-\omega_0$ is the detuning of the emitter from the target frequency, $\Gamma$ is the spontaneous emission linewidth, and $\tau$ is the delay between the control pulses). As a result, we can neglect dependence of the coupling constants $g_{\lambda,\bf k}$ on $\omega_{\bf k}$, and take $g_{\lambda,\bf k}=g$. Furthermore, the orientation of the momentum vector is irrelevant for our purposes, so we can enumerate the photon modes by their frequency, using the scalar index $k$. The multiplicity of photon states is taken into account via the density of the photon states $\rho_\omega=\omega^2/(2\pi^2 V)$, and, for the relevant frequencies within the narrow region near $\omega_0$, we can take the photon density of states $\rho_\omega$ as constant. 
This implies linear dependence of $\omega_k$ on index $k$ (as it should be, since $k$ enumerates the photon frequencies). 
Within RWA the zero frequency corresponds to $\omega_0$, so that $\omega_k$ denotes detuning between the frequency of the $k$-th mode and the target frequency $\omega_0$. Therefore, when considering  finite number $L$ of the photonic modes, we choose 
$\omega_k=-D+k\epsilon$, with $\epsilon=2D/(L-1)$; such choice corresponds to $\rho_\omega=1/\epsilon$. 
In reality the number of modes is very large, $L\to\infty$, since $L$ is proportional to the normalization volume $V$. This implies the scaling $g\propto L^{-1/2}$ and $\epsilon\propto L^{-1}$, as expected. 
Also, by including the RWA directly in the Hamiltonian and making the simplifications described above, we assume that $\omega_1$ is appropriately renormalized, and the non-Markovian effects in the electromagnetic bath can be neglected \cite{Milonni}.

Thus, we obtain the Hamiltonian
\begin{equation}
H = \sum_{k} \omega_k a^{\dagger}_{k}a_{k} + \frac{\Delta}{2} \sigma_z -i \sum_{k} g_{k} \left( a^{\dagger}_{k} \sigma^- - a_{k} \sigma^+ \right),
\end{equation}
where $\Delta$ is the detuning between the frequency of the emitter $\omega_1$ and the target frequency $\omega_0$, and $\omega_k$
is the detuning of the $k$-th mode from the target frequency $\omega_0$. 

In order to control the emission line of this system, we apply a sequence of pulses which periodically invert the state of the emitter, $\sigma_z\to -\sigma_z$. The pulses are spaced periodically in time, separated by an interval $\tau$, and have the following effect:
\begin{equation}
\sigma^{+}  \to \sigma^-, \; \sigma^- \to \sigma^{+}, \; \sigma_z \to -\sigma_z.
\end{equation}
If we consider time $t$ given by $t=n\tau + \delta$ with $n=0, 1, 2, \dots, M_p$, where $M_p$ is an even integer
and $\delta \in [0, \tau]$, then the Hamiltonian describing our two-level emitter coupled to a photon bath under the influence of the above pulse 
sequence at time $t$ is :
\begin{equation}
H(t)=\frac{\Delta}{2} (-1)^n \sigma_z + \sum_{k} \omega_k a^{\dagger}_{k}a_{k} + i \sum_{k} g_k \left\{ \xi_1 a^{\dagger}_{k} \sigma^- + \xi_2 a^{\dagger}_{k} \sigma^+ - \xi_1 a_k \sigma^+ - \xi_2 a_k \sigma^+ \right\}
\end{equation}
where we have introduced the filter functions $\xi_1(t)$ and $\xi_2(t)$ periodic in time with period $2\tau$ and defined by:
\begin{equation}
\xi_1(t) = \left\{ \begin{array}{l l}1 & \quad \mathrm{if} \quad t < \tau \\
0 & \quad \mathrm{if} \quad \tau < t < 2\tau \end{array} \right.
\end{equation}
and
\begin{equation}
\xi_2(t) = \left\{ \begin{array}{l l}0 & \quad \mathrm{if} \quad t < \tau \\
1 & \quad \mathrm{if} \quad \tau < t < 2\tau \end{array}\right.
\end{equation}
To evaluate the effect of this protocol, we will calculate the spectrum $N_k(t) = \langle a^{\dagger}_{k}(t)a_{k}(t) \rangle $.

\section{Equations of Motion and Recursive Relation}

The Heisenberg equations of motion for the emitter and photon operators are given by:
\begin{eqnarray}
\label{eq:eom2_1}
\dot a_k &=& -i \omega_k a_k + g_k \left( \xi_1 \sigma^- + \xi_2 \sigma^{+} \right) \\
\label{eq:eom2_2}
\dot \sigma^- &=&  -i (-1)^n \Delta \sigma^- + \sum_k g_k \xi_1 a_k \sigma_z - \sum_k g_k \xi_2 a^{\dagger}_k \sigma_z\\
\label{eq:eom2_3}
\dot \sigma_z &=& -2\sum_k g_k \left[ \xi_1 a^{\dagger}_k \sigma - \xi_2 a^{\dagger}_k \sigma^+ - \xi_2 a_k \sigma^- + \xi_1 a_k \sigma^{+} \right]
\end{eqnarray}
We want to calculate $N_k(t) =  \langle a^{\dagger}_k(t) a_k(t) \rangle$
with $t=M_p\tau + \delta$, $M_p$ a large even integer. After $M_p$ pulses, the equations of motion can be rewritten as:
\begin{equation}
\label{eq:eom3_1}
\dot a_k^{(M_p)} = -i \omega_k a_k^{(M_p)} + g_k \sigma^{- (M_p)}
\end{equation}
\begin{equation}
\label{eq:eom3_2}
\dot \sigma^{- (M_p)} =  -i \Delta \sigma^{- (M_p)} + \sum_k g_k a_k^{(M_p)} \sigma_z^{(M_p)} 
\end{equation}
\begin{equation}
\label{eq:eom3_3}
\dot \sigma_z = -2\sum_k g_k \left[ a^{\dagger (M_p)}_k \sigma^{- (M_p)} + a_k^{(M_p)} \sigma^{+ (M_p)} \right]
\end{equation}
where $O^{(M_p)}$ denotes the operator $O$ after $M_p$ pulses.

%\section{Recursive Relation}

Integrating Eq.~\ref{eq:eom3_1} between $M_p\tau$ and $t$ gives:
\begin{equation}
a_k^{(M_p)}(t) = a_k(M_p\tau) \mathrm{e}^{-i\omega_k \left( t - M_p\tau \right)} + g_k \int_{M_p\tau}^{t} \sigma^{- (M_p)}(t_1) \mathrm{e}^{-i\omega_k \left( t - t_1\right)} dt_1
\end{equation}
To obtain $a_k^{(M_p)}(M_p\tau)$ equation(\ref{eq:eom2_1}) is integrated between $(M_p-1)\tau$ and  $M_p\tau$ and gives:
\begin{equation}
a_k^{(M_p-1)}(M_p\tau) = a_k((M_p-1)\tau) \mathrm{e}^{-i\omega_k \tau} + g_k \int_{(M_p-1)\tau}^{(M_p\tau)} \sigma^{+ (M_p-1)}(t_1) \mathrm{e}^{-i\omega_k \left(M_p\tau - t_1\right)} dt_1
\end{equation}
By repeating this process recursively, we have:
\begin{eqnarray}
\label{eq:a_rec1}
a_k^{(M_p)}(t) &=& a_k(0) \mathrm{e}^{-i\omega_k t} + g_k \int_{M_p\tau}^{t} \sigma^{- (M_p)}(t_1) \mathrm{e}^{-i\omega_k \left( t - t_1 \right)} dt_1 \nonumber \\
&+& g_k \sum_{l=1,\; l \; odd}^{M_p-1} \int_{(M_p-l)\tau}^{(M_p-l+1)\tau} \sigma^{+ (M_p-l)}(t_1) \mathrm{e}^{-i\omega_k \left( t - t_1\right)} dt_1 \nonumber \\
&+& g_k \sum_{l=2, \; l \; even}^{M_p} \int_{(M_p-l)\tau}^{(M_p-l+1)\tau} \sigma^{- (M_p-l)}(t_1) \mathrm{e}^{-i\omega_k \left( t - t_1\right)} dt_1 
\end{eqnarray}

Introducing the filter functions $\xi_1(t)$ and $\xi_2(t)$, the
third term of Eq.~\ref{eq:a_rec1} can be expressed as follows. 
\begin{equation}
 g_k \sum_{l=0}^{M_p-1} \int_{l\tau}^{(l+1)\tau} \xi_2(t_1) \sigma^{+ (l)}(t_1) \mathrm{e}^{-i\omega_k \left( t - t_1\right)} dt_1.
\end{equation}
If we define $t_2$ such that $t_1 = t_2 + 2 l \tau$, $0<t_2< 2\tau$, since $\xi_2(t_1)=\xi_2(t_2)$, the above expression is equivalent to
\begin{equation}
g_k \sum_{l=0}^{M_p/2-1} \int_{0}^{2\tau} \xi_2(t_2) \sigma^{+ (l)}(t_2) \mathrm{e}^{-i\omega_k \left( t - t_2 -2l\tau \right)} dt_2.
\end{equation}
Applying the Markovian approximation to $\sigma^{+ (l)}(t_2) $, we can rewrite it as:
\begin{equation}
 g_k \sum_{l=0}^{M_p/2-1} p_k(\tau) \sigma^{+}(2l\tau) \mathrm{e}^{-i\omega_k \left( t - 2l\tau \right)}
\end{equation}
with 
\begin{eqnarray}
p_k(\tau) &=& \int_{0}^{2\tau} \xi_2(t_2) \mathrm{e}^{ i\left( \omega_k - \Delta \right) t_2} dt_2 \mathrm{e}^{i2\Delta \tau} \\
  &=& \frac{1}{i\left(\omega_k - \Delta \right)}\left[ \mathrm{e}^{i 2 \omega_k \tau} - \mathrm{e}^{i \left( \omega_k + \Delta \right) \tau} \right]
\label{eq:p_k}
\end{eqnarray}

Similarly, upon introducing the function $\xi_1(t)$ and invoking Markovian approximation on $\sigma^{- (l)}$,
the fourth term of Eq.~\ref{eq:a_rec1} can be rewritten as
\begin{equation}
g_k \sum_{l=0}^{M_p/2 -1} q_k(\tau) \sigma^-((2l+1)\tau) \mathrm{e}^{-i\omega_k \left( t - 2l\tau \right)}
\end{equation}
with
\begin{eqnarray}
q_k(\tau) &=& \int_{0}^{2\tau} \xi_1(t_2) \mathrm{e}^{ i\left( \omega_k - \Delta \right) t_2} dt_2 \mathrm{e}^{i\Delta \tau}\\
&=&\frac{1}{i\left(\omega_k - \Delta \right)}\left[ \mathrm{e}^{i  \omega_k \tau} - \mathrm{e}^{i \Delta \tau} \right]
\label{eq:q_k}
\end{eqnarray}

Altogether, we can rewrite Eq.~\ref{eq:a_rec1} as
\begin{eqnarray}
a_k^{(M_p)}(t) &=& a_k(0) \mathrm{e}^{-i\omega_k t} + g_k \int_{N\tau}^{t} \sigma^{- (M_p)}(t_1) \mathrm{e}^{-i\omega_k \left( t - t_1 \right)} dt_1 \nonumber \\
&+& g_k \sum_{l=1}^{M_p/2-1} p_k(\tau) \sigma^{+ }(2l\tau) \mathrm{e}^{-i\omega_k \left( t - 2l\tau \right)} \nonumber \\
&+& g_k \sum_{l=0}^{M_p/2-1}q_k(\tau) \sigma^-((2l+1)\tau) \mathrm{e}^{-i\omega_k \left( t - 2l\tau \right)} 
\end{eqnarray}
and $a^{\dagger}_k$ can be obtained by simply taking the adjoint of $a_k$:
\begin{eqnarray}
a^{\dagger (M_p)}_k(t) &=& a^{\dagger}_k(0) \mathrm{e}^{i\omega_k t} + g_k \int_{M_p\tau}^{t} \sigma^{+ (M_p)}(t_1) \mathrm{e}^{i\omega_k \left( t - t_1 \right)} dt_1 \nonumber \\
&+& g_k \sum_{l=1}^{M_p/2-1} p^*_k(\tau) \sigma^-(2l\tau) \mathrm{e}^{i\omega_k \left( t - 2l\tau \right)} \nonumber \\
&+& g_k \sum_{l=0}^{M_p/2-1} q^*_k(\tau) \sigma^{+}((2l+1)\tau) \mathrm{e}^{i\omega_k \left( t - 2l\tau \right)} 
\end{eqnarray}

\section{The Fluorescence Spectrum}

The number of photons in the $k$-th mode at time $t$ , $N_k(t)$ is given by:
\begin{equation}
N_k(t) = \langle a^{\dagger}_k(t) a_k(t) \rangle 
\end{equation}
We will evaluate this quantity by replacing $a^{\dagger}_k(t)$ and $a_k(t)$ with their expressions derived above. 
Since we are evaluating expectation values with respect to an initial state which has no photons (vacuum) and the emitter being in the excited state, we can drop terms involving $a^{\dagger}_k(0)$ and $a_k(0)$. Thus we obtain:
\begin{eqnarray}
N_k(t) &=& g_k^2 \int_{M_p\tau}^{t} \int_{M_p\tau}^{t} \sigma^{+ (M_p)}(t_1)\sigma^{- (M_p)}(t_2) \mathrm{e}^{-i\omega_k \left( t_1 - t_2 \right)} dt_1dt_2 \nonumber \\
&+& g_k^2 \; 2 \; \mathrm{Re} \; \left\{ \sum_{l=0}^{M_p/2-1} \int_{M_p\tau}^{t} p_k(\tau) \sigma^{+ (M_p)}(t_1) \sigma^{+ }(2l\tau) \mathrm{e}^{-i\omega_k \left( t _1- 2l\tau \right)} dt_1 \nonumber \right\}\\
&+& g_k^2 \; 2 \; \mathrm{Re} \;\left\{ \sum_{l=0}^{M_p/2-1} \int_{M_p\tau}^{t} q_k(\tau) \sigma^{+ (M_p)}(t_1) \sigma^-((2l+1)\tau) \mathrm{e}^{-i\omega_k \left( t _1- 2l\tau \right)} dt_1 \nonumber  \right\}\\
&+& g_k^2 \sum_{l,m=0}^{M_p/2-1}  p^*_k(\tau)p_k(\tau) \sigma^-(2l\tau) \sigma^{+ }(2m\tau) \mathrm{e}^{-i\omega_k \left( 2l\tau -2m\tau \right)} \nonumber \\
&+& g_k^2 \sum_{l,m=0}^{M_p/2-1} p^*_k(\tau) q_k(\tau)  \sigma^-(2l\tau) \sigma^-((2m+1)\tau) \mathrm{e}^{-i\omega_k \left( 2l\tau -2m\tau \right)} \nonumber \\
&+& g_k^2 \sum_{l,m=0}^{M_p/2-1} q^*_k(\tau) p_k(\tau) \sigma^{+}((2l+1)\tau) \sigma^{+ }(2m\tau) \mathrm{e}^{-i\omega_k \left( 2l\tau -2m\tau \right)} \nonumber \\
&+& g_k^2 \sum_{l,m=0}^{M_p/2-1} q^*_k(\tau) q_k(\tau) \sigma^{+}((2l+1)\tau) \sigma^-((2m+1)\tau)\mathrm{e}^{-i\omega_k \left( 2l\tau -2m\tau \right)} 
\label{eq:E_k}
\end{eqnarray}

To obtain $\sigma^{\left( + , - \right)}(n\tau)$, we integrate the corresponding equations of motion iteratively.  Consider Eq.~\ref{eq:eom2_2}:
\begin{equation}
\dot \sigma^{- (n)} =  -i (-1)^n \Delta \sigma^- + \sum_k g_k \xi_1 a_k \sigma_z - \sum_k g_k \xi_2 a^{\dagger}_k \sigma_z
\end{equation}
For $n=0$, before the first pulse, it reads:
\begin{equation}
\label{eq:sigma_0_2}
\dot \sigma^- =  -i \Delta \sigma^- + \sum_k g_k a_k \sigma_z.
\end{equation}
Plugging in the corresponding $a_k(t)$ gives
\begin{equation}
\dot \sigma^- =  -i \Delta \sigma^- + \sum_k g_k a_k(0) \mathrm{e}^{-i\omega_k t}\sigma_z(t) + \sum_k g_k^2 \int_0^t \sigma^-(t_1) \sigma_z(t) \mathrm{e}^{-i\omega_k \left( t -t_1 \right)} 
\end{equation}
Upon applying the Markovian approximation to $\sigma^-$ on the right-hand side, this can be rewritten as:
\begin{equation}
\dot \sigma^- = -i \Delta \sigma^- + \sum_k g_k^2 r^0_k(t) \sigma^-(t) \sigma_z(t) + \sum_k g_k a_k(0) \mathrm{e}^{-i\omega_k t}\sigma_z(t).
\end{equation}
Dropping the term with $a_k(0)$ which will involve vacuum fluctuations, using the identity $\sigma^-(t) \sigma_z(t) = \sigma^-(t)$ and 
defining $\beta^{(0)}(t) = \sum_k g_k^2 r^{(0)}_k(t)$ with  $r^{(0)}_k(t) = \int_0^t \mathrm{e}^{-i(\omega_k - \Delta)(t-t_1)}dt_1$,  we get
\begin{equation}
\label{eq:ZeroPulseEOM}
\dot \sigma^- =  -i \left( \Delta +i \beta^{(0)}(t) \right) \sigma^-
\end{equation}
so that 
\begin{equation}
\label{eq:ZeroPulseSigma}
\sigma^-(\tau) = \mathrm{e}^{-i \int_0^{\tau} \left( \Delta + i\beta^{(0)}(s) \right) ds} \sigma^-(0)
\end{equation}
with
\begin{equation}
\label{eq:ZeroPulseGamma}
\beta^{(0)}(t) = \sum_k g_k^2 \frac{1-\mathrm{e}^{-i \left( \omega_k -\Delta \right) t }}{i \left( \omega_k -\Delta \right)}.
\end{equation}
This leads to
\begin{equation}
\sigma^-(\tau) = \mathrm{e}^{-i \left( \Delta \tau + i \gamma^{(0)}(\tau) \right)} \sigma^-(0)
\end{equation}
with:
\begin{equation}
\gamma^{(0)}(\tau) =  \sum_k g_k^2 \frac{\tau }{i \left( \omega_k -\Delta \right)} - \sum_k g_k^2 \frac{\mathrm{e}^{-i \left( \omega_k -\Delta \right)\tau} -1}{\left( \omega_k -\Delta \right)^2}. 
\end{equation}
For $n=1$, after the first pulse, the equation of motion reads:
\begin{equation}
\label{eq:sigma_1}
\dot \sigma^- =  i \Delta \sigma^- - \sum_k g_k a^{\dagger}_k \sigma_z.
\end{equation}
We plug in the expression of $a^{\dagger}_k$ and only keep terms up to second order in $g_k$. This approximation is justified by the fact 
that we have a large number of photon modes and as a result, each $g_k$ can be considered small compared to the other energy scales in the problem.
We obtain
\begin{equation}
\dot \sigma^- =  i \Delta \sigma^- - \beta^{(1)} \sigma^-(t) 
\end{equation}
with $\beta^{(1)}(t)= \sum_k g_k^2 r^{(1) *}_k(t)$ where $r^{(1)}_k(t) = \int_{\tau}^t \mathrm{e}^{-i(\omega_k - \Delta)(t-t_1)}dt_1$. This leads to 
\begin{equation}
\sigma^- (2\tau) = \mathrm{e}^{i \int_{\tau}^{2\tau} \left( \Delta + i\beta^{(1)}(s) \right) ds} \sigma^-(\tau) 
\end{equation}
with
\begin{equation}
\beta^{(1)}(t)=\sum_k g_k^2 \left[ \frac{1-\mathrm{e}^{-i \left(\omega_k -\Delta\right)(t - \tau)}}{i \left(\omega_k -\Delta\right)} \right]^*.
\end{equation}
From this we get 
\begin{equation}
 \sigma^- (2\tau) = \mathrm{e}^{i \left( \Delta \tau + i \gamma^{(1)} \right)} \sigma^- (\tau) = \mathrm{e}^{i \left( \Delta \tau + i \gamma^{(0)*} \right)} \sigma^- (\tau).
\end{equation}

Carrying this iterative process out to arbitrary $n$ will yield:
\begin{equation}
\label{eq:sigma_n_1}
\sigma^-(n\tau) = \mathrm{e}^{\left\{ \gamma^{(0)} - \gamma^{(1)} +\gamma^{(2)}- \gamma^{(3)} \cdots -i\xi_1(n\tau) \Delta\right\}\tau} \sigma^-(0).
\end{equation}
From the observation
\begin{eqnarray}
\gamma^{(1)} &=& \gamma^{(0) *} \nonumber \\
\gamma^{(2)} &=& \gamma^{(0)} \nonumber \\
\gamma^{(3)} &=& \gamma^{(0) *} \nonumber \\
&\vdots&  \nonumber \\
\gamma^{(2n)} &=& \gamma^{(0)} \nonumber \\
\gamma^{(2n+1)} &=& \gamma^{(0) *}
\end{eqnarray}
Eq.~\ref{eq:sigma_n_1} becomes
\begin{eqnarray}
\sigma^-(2n\tau) &=& \mathrm{e}^{ i \left\{ 2  n \; \mathrm{Im} \gamma^{(0)}  \right\}} \sigma^-(0)\\
\sigma^-((2n+1)\tau)&=& \mathrm{e}^{ \left\{ i\; 2 n \; \mathrm{Im} \gamma^{(0)} + \gamma^{(0)} -i \Delta \tau \right\}} \sigma^-(0)
\end{eqnarray}
with
\begin{equation}
\label{eq:gamma_0}
\gamma^{(0)}(\tau) =  \sum_k g_k^2 \frac{\tau }{i \left( \omega_k -\Delta \right)} - \sum_k g_k^2 \frac{\mathrm{e}^{-i \left( \omega_k -\Delta \right)\tau} -1}{\left( \omega_k -\Delta \right)^2} 
\end{equation}

After the $M_p^{th}$ pulse, the process described above yields:
\begin{equation}
\sigma^{- (M_p)}(t) =\sigma^{- (M_p)} \left( M_p\tau \right) \mathrm{e}^{-i\int^t_{M_p\tau}( \Delta + i \beta^{(M_p)}(t_1)) dt_1} 
\end{equation}
Similarly,
\begin{equation}
\sigma^{+ (M_p)}(t) =\sigma^{+ (M_p)} \left( N\tau \right) \mathrm{e}^{i\int^t_{M_p\tau}( \Delta - i \beta^{(M_p)*}(t_1)) dt_1 }
\end{equation}
with
\begin{eqnarray}
\beta^{(M_p)}(t) = \sum_k g_k^2  \frac{1 - \mathrm{e}^{-i(\omega_k - \Delta) \left( t -M_p\tau \right)}}{ -i(\omega_k - \Delta)}
\end{eqnarray}
Putting it all together, we will evaluate the right-hand side of Eq.~\ref{eq:E_k} term by term so that 
$N_k(t)=\sum_{i=1}^7 N^k_i(t)$. The first term is:
\begin{equation}
N^k_1 = g_k^2 \; \frac{1}{2} \langle \sigma_z(0) + I \rangle \left| \Psi_{\omega_k}(t) \right|^2
\end{equation}
with
\begin{equation}
\Psi_{\omega_k}(t) =\int_{0}^{t-M_p\tau} \mathrm{exp} \left\{ -i \Delta t_1 + \sum_{k'} g_{k'}^2  \left[ \frac{t_1}{i(\omega_{k'} - \Delta)} 
- \frac{\mathrm{e}^{-i(\omega_{k'} - \Delta) t_1 } -1}{ (\omega_{k'} - \Delta)^2} \right] \right\} \mathrm{e}^{i\omega_k t_1 } dt_1
\end{equation}
 
The second, the fifth and the sixth terms  on the right-hand side of Eq.~\ref{eq:E_k} will vanish since they are proportional to 
$\sigma^{+} \sigma^{+}$ or $\sigma^- \sigma^-$ equal-time products.
The third term is:
\begin{eqnarray}
N^k_3 &=& g_k^2 \; 2\; \mathrm{Re} \; \Bigg\{ q_k(\tau) \frac{1}{2} \langle \sigma_z(0) + I \rangle\left( \mathrm{e}^{ \left\{ i \; M_p \; \mathrm{Im} \; \gamma^{(0)} \right\} } \right)^*  \nonumber \\
&\times& \sum_{l=0}^{M_p/2 - 1}\mathrm{e}^{ i \left\{ 2l\; \mathrm{Im} \gamma^{(0)} + \gamma^{(0)} -i\Delta \tau  \right\}}  \mathrm{e}^{-i\omega_k \left(M_p\tau - 2l\tau\right)} \Psi_{\omega_k}(t)^* \Bigg\}.
\end{eqnarray}

The fourth term is:
\begin{eqnarray}
N^k_4 &=& g_k^2 \; p_k^*(\tau) p_k(\tau) \; \langle I -\frac{1}{2}\left( \sigma_z(0) + I \right) \rangle \nonumber \\
&\times& \sum_{l, m=0}^{M_p/2 -1}  \mathrm{e}^{ i \left\{ 2  l \; \mathrm{Im} \gamma^{(0)}  \right\}} \left( \mathrm{e}^{ i \left\{ 2  m \; \mathrm{Im} \gamma^{(0)}  \right\}}\right)^* \mathrm{e}^{ i \omega_k \left( 2m -2l \right) \tau } 
\end{eqnarray}
and finally the seventh term is:
\begin{eqnarray}
N^k_7 &=& g_k^2 \; \frac{1}{2} \langle \sigma_z(0) + I \rangle q_k^*(\tau) q_k(\tau) \nonumber \\
&\times& \sum_{l, m =0}^{M_p/2 -1} \mathrm{e}^{ i \omega_k \left( 2m -2l \right) \tau } \left( \mathrm{e}^{ \left\{ i\; 2 l \; \mathrm{Im} \gamma^{(0)} + \gamma^{(0)} -i \Delta \tau \right\}}\right)^* \mathrm{e}^{ \left\{ i\; 2 m \; \mathrm{Im} \gamma^{(0)} + \gamma^{(0)} -i \Delta \tau \right\}} 
\end{eqnarray}
with $p_k(\tau)$ and $q_k(\tau)$ given by Eq.~\ref{eq:p_k} and Eq.~\ref{eq:q_k} and $\gamma^{(0)}$ by Eq.~\ref{eq:gamma_0}.

Our results are mapped onto the frequency space. To evaluate the sums over $k$, we transform them into integrals in frequency, taking into account constant density of modes $\rho_{\omega} = 1/\epsilon$ obtained from the dispersion relation $\omega_k = -D_{\omega} + k \epsilon$ where $\epsilon=2D_{\omega}/(L-1)$. The expressions above are evaluated numerically; for most calculations we use $L=151$ and $L=201$ in order to compare them with the results of the tDMRG numerical simulations. We normalize $\epsilon$ and $g$ such that the width at half-maximum of the spectrum in the absence of pulses is $\Gamma=2$. We calculate the integrals in an interval from $-D$ to $+D$, choosing $D=20$ and $D=30$ for our calculations. 

As can be inferred from Eq.\ref{eq:ZeroPulseEOM}, $\Gamma$ is given by the real part of $\beta^{(0)}(t \to \infty )$. The transformation of the summation in Eq.\ref{eq:ZeroPulseGamma} into an integral introduces an error of 
order $\mathcal{O}(g^2 \epsilon/(D_{\omega}-\Delta), g^2 \epsilon/(D_{\omega}+\Delta))$.
An asymptotic expansion of the integral is obtained as:
\begin{eqnarray}
 \beta^{(0)}(t) &=& g^2 \; \pi + i g^2 \; \mathrm{ln} \frac{D_{\omega} + \Delta}{D_{\omega} - \Delta} - g^2 \; \mathrm{e}^{i( D_{\omega} - \Delta)t}
 \left[ \frac{1}{(D_{\omega} - \Delta)t} - \frac{i}{(D_{\omega} - \Delta)^2 t^2} \;\; \cdots \;\; \right] \nonumber \\
   & - & g^2 \; \mathrm{e}^{-i( D_{\omega} - \Delta)t} 
   \left[ \frac{1}{(D_{\omega} - \Delta)t} + \frac{i}{(D_{\omega} - \Delta)^2 t^2} \;\; \cdots \;\; \right] \nonumber \\
   & - & g^2 \; \mathrm{e}^{i( D_{\omega} + \Delta) t} 
   \left[ \frac{1}{(D_{\omega} + \Delta)t} - \frac{i}{(D_{\omega} + \Delta)^2 t^2} \;\; \cdots \;\; \right] \nonumber \\
   & + & g^2 \; \mathrm{e}^{-i( D_{\omega} - \Delta)t} 
   \left[ -\frac{1}{(D_{\omega} - \Delta)t} - \frac{i}{(D_{\omega} - \Delta)^2 t^2} \;\; \cdots \;\; \right]
\end{eqnarray}
Thus, our choice of $g$ introduces an error of order $\mathcal{O}(g^2/(D_{\omega}-\Delta)t, g^2/(D_{\omega}+\Delta)t )$. 
Altogether, corrections to our results at time $t$ are of order $\mathcal{O}(g^2 \epsilon/(D_{\omega}-\Delta), g^2/(D_{\omega}-\Delta)t)$.

\section{Robustness of the protocol with respect to small imperfections}

\begin{figure}
\includegraphics[width=8cm]
{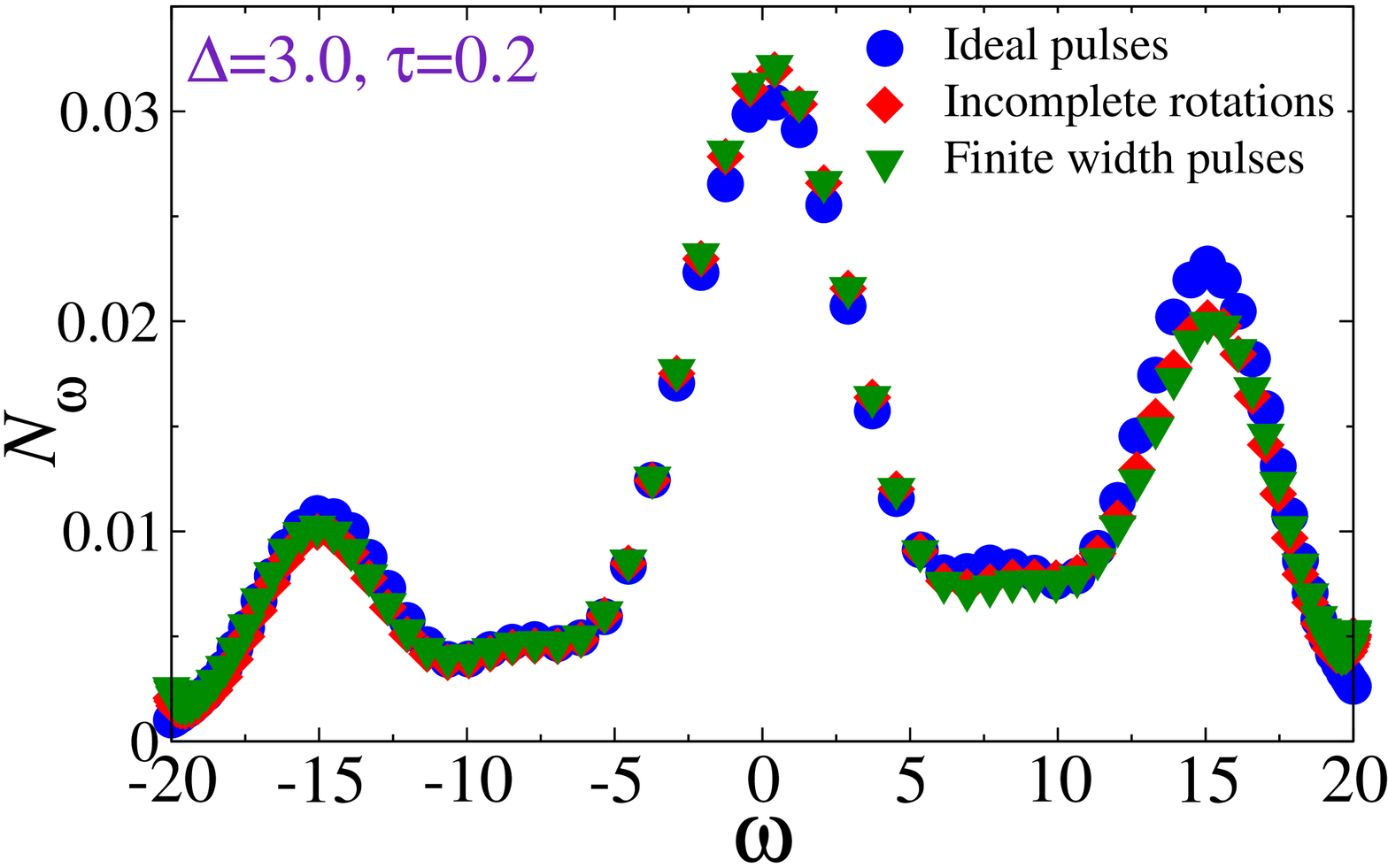}  
\caption{(Color online) Emission spectra $N_\omega$ for $\tau=0.2$ and $\Delta=3.0$ after 6 pulses. The results for ideal, instant 180$^\circ$, pulses (blue circles) are compared with those for imperfect pulses: red diamonds --- instant pulses with incomplete rotations, 175$^\circ$ instead of ideal 180$^\circ$; green triangles --- pulses of finite width $t_p=0.05$. All three spectra practically coincide, showing robustness of the protocol.} 
\label{fig:nonIdealPulses}
\end{figure}

In realistic experiments, the pulses are never ideal: they have finite width, and the rotation angle may slightly deviate from exact 180$^\circ$. We explicitly tested robustness of our approach with respect to these two most typical experimental non-idealities. The results of the corresponding tDMRG simulations in Fig.~\ref{fig:nonIdealPulses} are compared with the results for ideal (instantaneous 180$^\circ$) pulses. We find that a moderate 5$^\circ$ error in the rotation angle does not affect efficiency of the control. In the same way, pulses as wide as $t_p=0.05$ (which is 1/4 of the inter-pulse distance $\tau$) remain as efficient as ideal pulses. Thus, the protocol can be used in realistic systems, such as NV centers, with realistically achievable parameters of the pulses (Rabi driving frequency, pulse width, and experimental jitters).

%\begin{figure}[htbp]
%\includegraphics*[width=8.0cm]{simultnImperfFiniteWidthDelta3p0tau0p2_Np6.pdf}  

\section{Additional results}

\begin{figure}
\includegraphics[width=9.0cm]{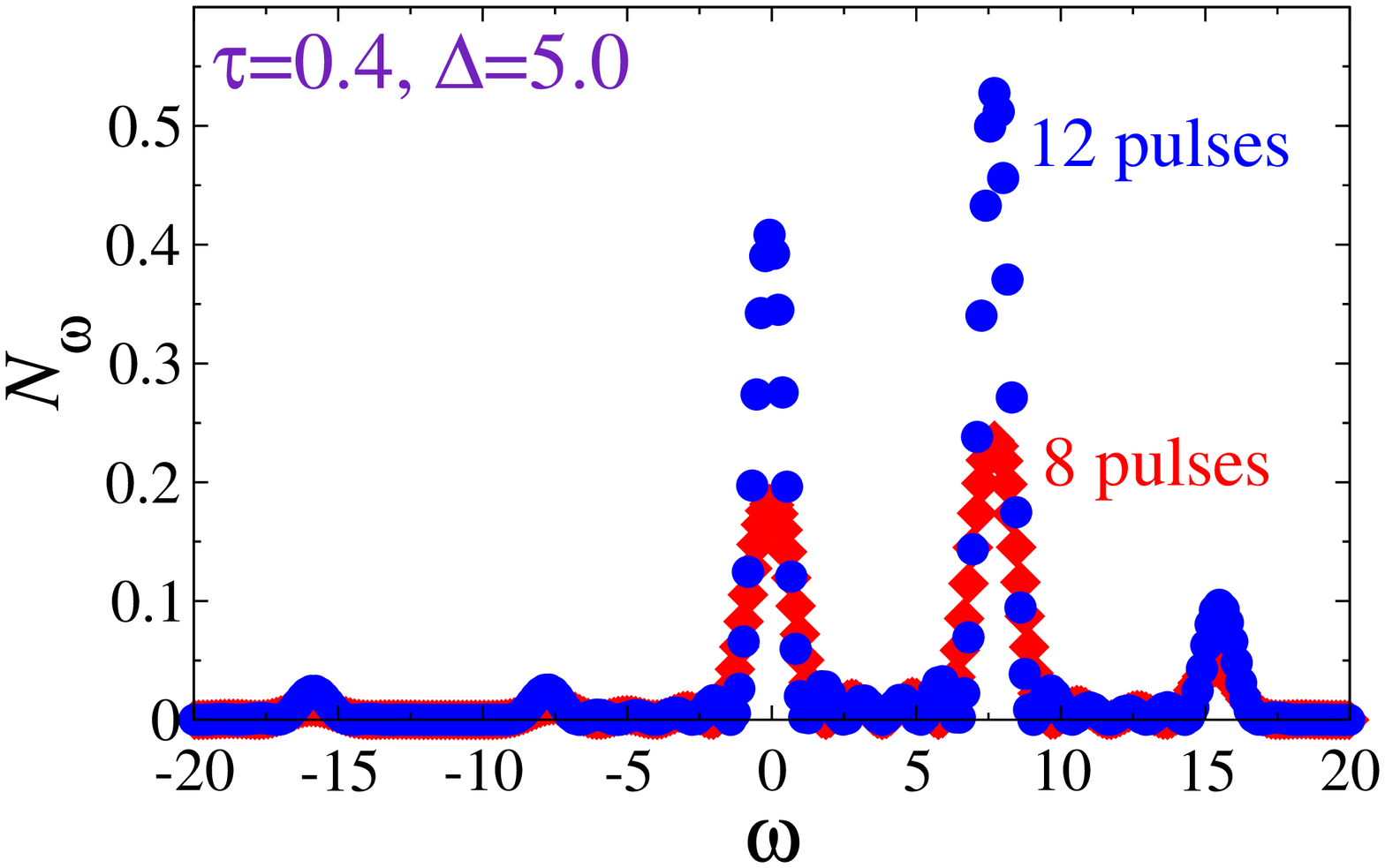} 
\caption{(Color online) Evolution in time of the emission profile. Shown after  $8$ (red diamonds) and $12$ (blue circles)  pulses for $\tau=0.4$ and $\Delta=5.0$. The negative-frequency satellite peak is suppressed and more of the spectral weight is contained in the positive-frequency satellite peak closest to $\omega_1$.} 
\label{fig:satelitePositiveFreq}
\end{figure}

\begin{figure}
\includegraphics[width=9.0cm]{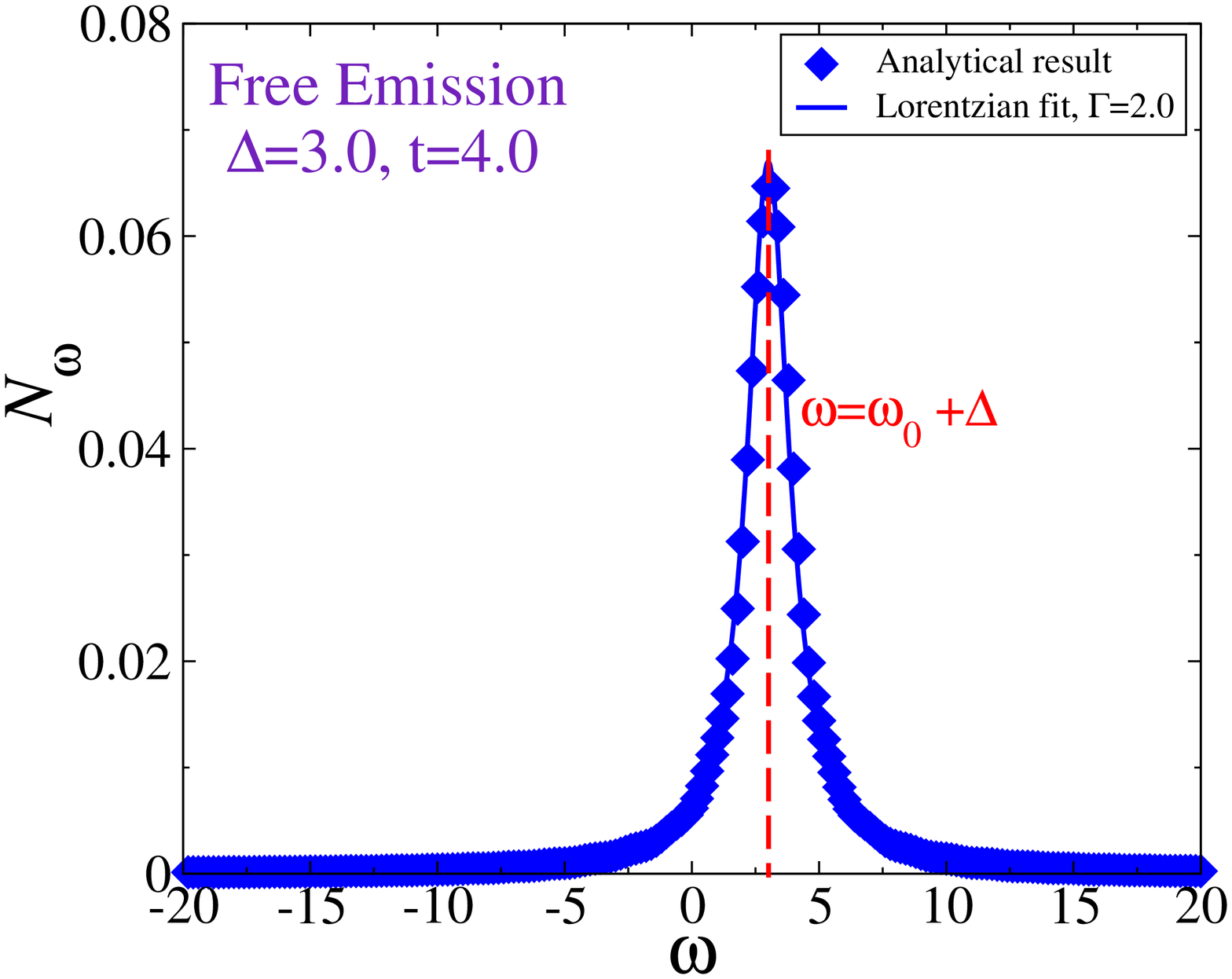} 
\caption{(Color online) $N_\omega$ for a control-free emission after time $t=4.0$ for $\Delta=3.0$. The symbols represent analytical results, while the solid line represents a Lorentzian centered at $\omega_1$ with the width $\Gamma=2.0$.} 
\label{fig:freeEmission}
\end{figure}

Increasing both $\tau$ and $\Delta$ can eventually suppress the negative-frequency satellite peak, and lead to a situation where the positive-frequency 
satellite peak (i.e.\ the peak closest to the emitter's original frequency $\omega_1$) contains most of the spectral weight, as seen in Fig.~\ref{fig:satelitePositiveFreq}. This is to be expected as the situation with $\Delta\gg \tau^{-1}$ comes closer the limit of no control. Contrasting this with the free emission spectrum after time $t = 4.0$ (Fig.~\ref{fig:freeEmission}), it is worth noting that as this limit is approached, the protocol is still preferable to the free emission: a significant fraction of the emission is still happening at the target frequency.

\section{tDMRG simulations}

In order to efficiently carry out the tDMRG simulations with periodic boundary conditions we implemented the change of basis described in Ref.\onlinecite{Feiguin2011}. Since the bosonic bath is modeled by a non-interacting one dimensional chain, one can perform a folding transformation in real space by rotating to a single-particle basis defined by the operators

\[
a_{i,\pm} = \frac{1}{\sqrt{2}}\left[ a_{iL} \pm a_{iR} \right], 
\]
where the chain has been split into left ($L$) and right ($R$) sites relative to the position of the two-level emitter. As a consequence of the transformation, the emitter will couple only to the symmetric (+) channel, and decouple completely from the antisymmetric one. However, one should notice that the complex hopping is equivalent to a magnetic flux threading the ring, and therefore, the reflection symmetry is broken. Consequently the (+) and (-) channels will couple as described in Fig.1(c) of Ref.\onlinecite{Feiguin2011}. 

The main effect of the transformation is to map the problem onto an equivalent one with open boundary conditions, while keeping the Hamiltonian local. This represents a dramatic reduction of the entanglement and the computational cost of the time-evolution. Moreover, open boundary conditions enable us to use a Suzuki-Trotter decomposition of the evolution operator. In our simulations, the number of bosons never grows larger than the number of pulses, which allows us to truncate the size of the local basis, and the number of DMRG states needed is never larger than 100, while the truncation error is always smaller than $10^{-9}$. 
%As a matter of fact, most of the computational cost consists of calculating all the correlations for the momentum distribution.

\section{Analysis based on the master equation for the emitter}

The purpose of this section is three-fold. First, we demonstrate that the analysis based on the master equation for the emitter produces correct description of the system, and in particular, of the spectrum of the emitted light. Second, the master equation approach provides a more formal way to elucidate the basic ideas underlying the proposed method for controlling the emission spectrum, and demonstrate qualitatively why the method works. Third, we calculate explicitly the rate of the coincidence counts for the two-photon interference (TPI) experiments, and explicitly show that the method for control of the emission spectrum indeed greatly improves the indistinguishability of the photons emitted by two different emitters, with different detunings $\Delta_1$ and $\Delta_2$.

\subsection{Master equation for the emitter and the correlation functions}

Under the standard set of assumptions, which are usually satisfied in typical experiments, and which are also approximately satisfied in the calculations above (both analytical and numerical), the equations of motion for the reduced density matrix $\rho$ of the two-level emitter can be obtained, see Ref.~\onlinecite{CohenTann} for details. The assumptions are: (i) applicability of the Markov approximation for the emitter density matrix, which implies that the electromagnetic radiation bath is not strongly perturbed by emission, and/or returns to its equilibrium state quickly, and (ii) applicability of the rotating-wave approximation, discussed in the main text. In particular, it is implied that the number of modes $L\to\infty$, the total photon spectral range $D\to\infty$, and the density of photon states is linear and equal to $2D/L$ in the relevant spectral region.

Representing the emitter's density matrix as
\begin{equation}
\rho = \rho_{ee}|e\rangle\langle e| + \rho_{gg}|g\rangle\langle g| + \rho_{eg}|e\rangle\langle g| +\rho_{ge}|g\rangle\langle e|,
\end{equation}
the equations of motion between the pulses can be obtained, and written (in the frame rotating with the target frequency $\omega_0$) in the form
\begin{eqnarray}
\dot\rho_{ee} &=& -\Gamma\rho_{ee}\\
\dot\rho_{gg} &=& \Gamma\rho_{gg}\\
\dot\rho_{ge} &=& i\Delta\rho_{ge} - \Gamma\rho_{ge}/2\\
\dot\rho_{eg} &=& -i\Delta\rho_{eg} - \Gamma\rho_{eg}/2.
\label{rhoevol}
\end{eqnarray}
The initial conditions (no emitted photons, emitter in the excited state) are $\rho_{ee}=1$, $\rho_{gg}=\rho_{ge}=\rho_{eg}=0$.
Ideal (instantaneous, 180$^\circ$) pulses are applied to the emitter at time instants $t=k\tau$ (with $k=1,2,\dots$), and they transform the density matrix as follows:
\begin{equation}
\rho(k\tau+0) = \sigma_x \rho(k\tau-0) \sigma_x
\end{equation}
where $\sigma_x$ is the Pauli $x$--matrix, while $\rho(k\tau-0)$ and $\rho(k\tau+0)$ are the emitter's density matrix before and after the pulse, respectively.

{\bf It is important to note here} that the transformation of $\rho$ under the action of the pulse has such a simple form only in the rotating frame chosen above. In a different frame (laboratory frame, or some other rotating frame) the transformation law of the density matrix would be rather complex.

Both the emission spectrum and the photon indistinguishability are determined by the emitter's correlation function 
\begin{equation}
\label{cf}
\phi(t,\theta)=\langle\sigma^+(t+\theta)\sigma^-(t)\rangle,
\end{equation}
where the angle brackets denote quantum-mechanical average (trace of the operator over the emitter's reduced density matrix).
Indeed, within the approximations outlined above, the photon creation operator at the space point ${\mathbf r}_s$ and time $t_s$ is 
\begin{equation}
a^\dag({\mathbf r}_s, t_s) = A \sigma^+(t_s-|{\mathbf r}_s|/c)
\label{aVSsigma}
\end{equation}
where $c$ is the speed of light and $A$ is a proporionality coefficient (irrelevant for our purposes), see Ref.~\onlinecite{ScullyZubairy}. As a result, both emission spectrum and the photon indistinguishability, being governed by the averages like $\langle a^\dag({\mathbf r}_s, t_s) a({\mathbf r}_s, t_s+\theta)\rangle$, can be expressed via $\phi(t,\theta)$, see Refs.~\onlinecite{CohenTann,ScullyZubairy}.

\begin{figure}
\includegraphics[width=9.0cm]{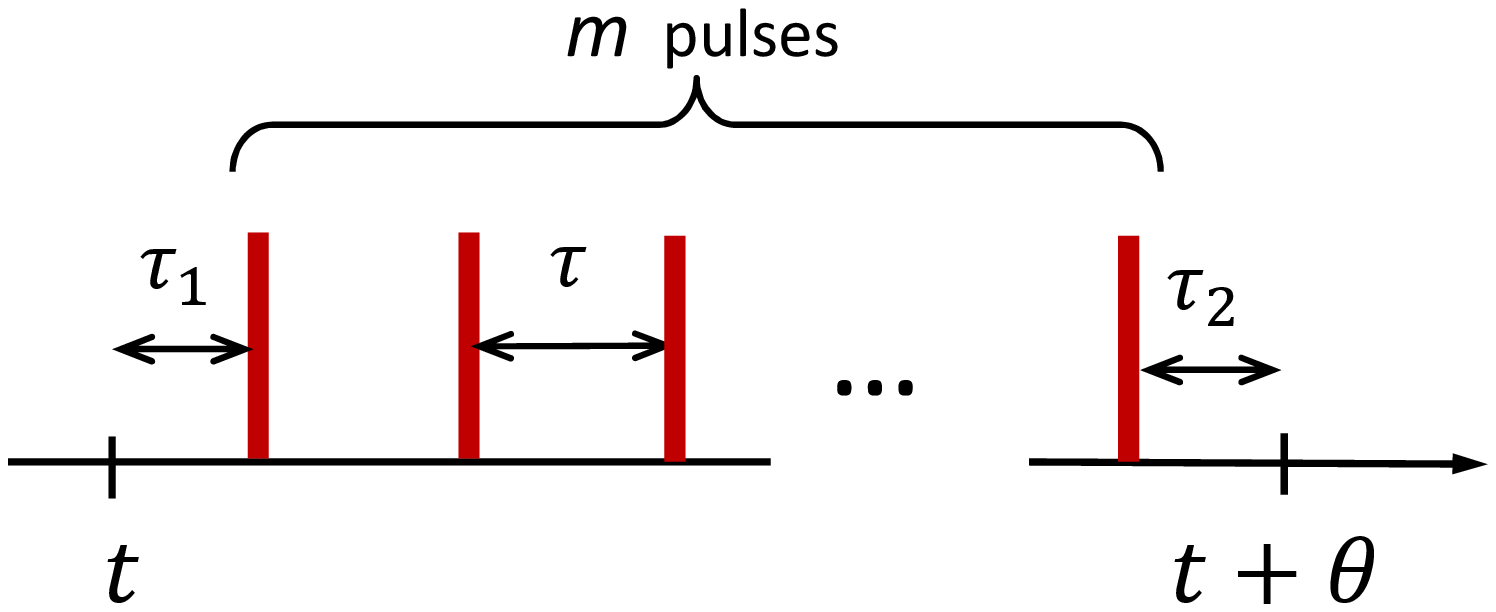} 
\caption{(Color online) Schematic picture of the mutual positions of the time instants $t$ and $t+\theta$ with respect to the pulses.} 
\label{SIfigPulses}
\end{figure}

With the master equation, we can find the correlation function $\phi(t,\theta)$ using the method described in Ref.~\onlinecite{Mollow} (in essense, simple version of the quantum regression theorem). It has the general form
\begin{equation}
\label{ffunc}
\phi(t,\theta)=\rho_e(t) f(t,\theta).
\end{equation}
Here $\rho_e(t)$ is the population of the $|e\rangle$ level at time $t$, and $f(t,\theta)$ is the function directly characterizing coherence and spectral properties of the emitter. 

To characterize the correlation function, we consider the two time instants, $t$ and $t+\theta$, as shown in Fig.~\ref{SIfigPulses}. I.e.\ we represent $\theta=\tau_1+\tau_2+(m-1)\tau$ where $m$ is the number of pulses separating $t$ and $t+\theta$ and $\tau_{1,2}\in[0,\tau)$; we also represent $t$ in a similar way, as $t=M\tau + (\tau-\tau_1)$ (so that $M$ is the number of pulses separating $t$ from the origin). The quantity $\rho_e(t)$ then can be found as
\begin{equation}
\rho_e(t)=\frac{1-(-1)^{M+1}e^{-\Gamma\tau (M+1)}}{1+e^{-\Gamma\tau}}\ \exp{[-\Gamma(\tau-\tau_1)]}.
\end{equation}
The function $f(t,\theta)$ has more complex form. When both $t$ and $t+\theta$ belong to the same interval between two subsequent pulses (i.e.\ $m=0$), we have
\begin{equation}
f(t,\theta)=\exp{(-\Gamma\theta/2)}\exp{(i\theta\Delta)}.
\end{equation}
When $t$ and $t+\theta$ are separated by the odd number of pulses 
\begin{equation}
f(t,\theta)=0,
\end{equation}
and if the number of pulses $m$ between these two instants is even, then
\begin{equation}
f(t,\theta)=\exp{(-\Gamma\theta/2)} \ \exp{[i\Delta(\tau_1+\tau_2-\tau)]}.
\end{equation}

This form of the correlation function {\bf provides a more formal qualitative description of the basic idea of our approach}. Without pulses, the correlation function of the emitter would include an oscillating factor $\exp{[i\theta\Delta]}$. Since the spectrum is governed by the Fourier transform of the correlation function (see below for detail), it is the oscillation that determines the peak in the emission spectrum at the frequency $\Delta$. In contrast, in the presence of pulses, the oscillations are constantly interrupted and reversed by the pulses. As long as $\tau\Delta$ is small, the function $f(t,\theta)$ is kept almost constant during even pulse intervals (apart from slow spontaneous decay with the rate $\Gamma$), and jumps to zero during odd pulse intervals. The oscillations at the frequency $\Delta$ are practically wiped out by the pulses: the second factor in the equation above always stays close to 1. The exact value of $\Delta$ becomes irrelevant, and the correlation function behaves as if $\Delta$ was always zero, i.e.\ as if the emitter always had the target frequency, without any spectral diffusion. 

This behavior determines both the emission spectrum and the enhanced indistinguishability of the photons. Both features are explicitly shown below.

\subsection{Emission spectrum analysis based on emitter's correlation function}

The spectrum analyzer (narrow-band detector) can be modeled as a two-level absorber with a very sharp transition frequency $\omega$, see Ref.~\cite{ScullyZubairy}. The excitation probability of the detector (placed at the space point $\mathbf r$) during the period of time $t\in[0,T]$ 
\begin{equation}
P(\omega,T)=\left\langle\left|\int_0^T a({\mathbf r},t)\exp{[i\omega t]} dt\right|^2\right\rangle
\end{equation}
is determined by the emitter's correlation function, as
\begin{equation}
P(\omega,T)=A^2 \int_0^T dt \int_0^T ds\  
\langle \sigma^+(t)\sigma^-(s)\rangle\exp{[-i\omega(t-s)]},
\end{equation}
or, taking into account that $\langle \sigma^+(t)\sigma^-(s)\rangle=\langle \sigma^+(s)\sigma^-(t)\rangle^*$,
we can write it as
\begin{equation}
P(\omega,T)=2\ A^2 {\rm Re}\ \int_0^T dt \int_0^{T-t} d\theta\,
\langle \sigma^+(t+\theta)\sigma^-(t)\rangle\exp{(-i\omega\theta)}.
\label{pintegr}
\end{equation}

\begin{figure}
\includegraphics[width=9.0cm]{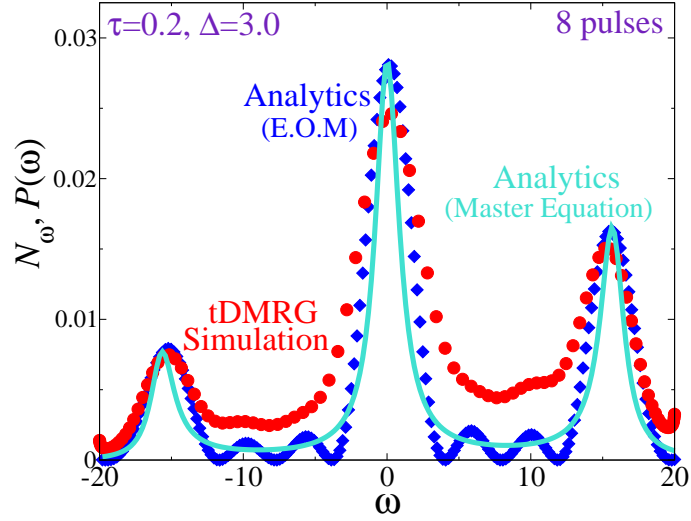} 
\caption{(Color online) Comparison between all three approaches used in this work: tDMRG simulations (red dots), analytics based on joint equations of motion for the emitter and the electromagnetic field (blue diamonds), and the analysis based on master equation (cyan line), for $\Delta=3.0$, $\tau=0.2$, after 8 pulses. To plot the master equation spectrum, the long-time value $P(\omega)$ given by Eq.~\ref{longtimeP} was used, and the overall amplitude of $P(\omega)$ was scaled to coincide with the height $N_{\omega=0}$ of the central peak obtained from the equations of motion.} 
\label{SIfigAllResCompare}
\end{figure}

The value $P(\omega,T)$ can be evaluated explicitly; its general form is quite complex, but can be significantly simplified for the experimentally relevant case of long times $T\gg t_0=1/\Gamma$. For simplicity, we can assume $T=2K\tau$, where $K\gg 1$ is a large positive integer. The result is:
\begin{eqnarray}
P(\omega)&=& 2 A^2 {\rm Re}\ \frac{1}{\gamma_0 (1+e^{-\Gamma\tau})}\left[\left(\frac{1-e^{-\Gamma\tau}}{\Gamma} - e^{-\gamma_0\tau} \frac{e^{\gamma_4\tau}-1}{\gamma_4}\right)\left(K+\frac{e^{-\Gamma\tau}}{1+e^{-\Gamma\tau}}\right)\right. \\ \nonumber
&+& \left.\frac{e^{\gamma_4\tau}-1}{\gamma_4} \left(1-e^{-\gamma_0\tau}\right)\frac{e^{-2\gamma_3\tau}}{1-e^{-2\gamma_3\tau}} \left(K+\frac{e^{-\Gamma\tau}}{1+e^{-\Gamma\tau}} - \frac{e^{-2\gamma_3\tau}}{1-e^{-2\gamma_3\tau}}\right)\right],
\label{longtimeP}
\end{eqnarray}
where $\gamma_0=i(\omega-\Delta)+\Gamma/2$, $\gamma_3=i\omega + \Gamma/2$, and $\gamma_4=i(\omega-\Delta)-\Gamma/2$.

This expression can be simplified in the case of small inter-pulse delays, where $\tau\Delta\ll 1$ and $\tau\Gamma\ll 1$. Considering the vicinity of the target frequency, where $\omega\tau$ is also a small quantity, and leaving only the largest terms, we obtain
\begin{equation}
P(\omega)=A^2\ \frac{K\tau\Gamma}{4}\ \frac{1}{\omega^2 + (\Gamma/2)^2}
\end{equation}
i.e.\ the Lorentzian line with the emitter's natural width, but now centered at the target frequency, not the original frequency $\Delta$, with the amplitude proportional to the total time $T$. 
Similar analysis can be performed for other regions of the spectrum (although with more caution, since $\omega\tau$ is not necessarily small already), to confirm the spectrum general structure and appearance of the side peaks.

Fig.~\ref{SIfigAllResCompare} illustrates comparison of the spectra obtained from the master equation analysis with the results produced by tDMRG simulations and by the analytics based on the joint equations of motion for the electromagnetic field and the emitter. It is seen that the master equation approach produces correct description of the spectra, and agrees very well with the two other approaches used in this work.

\begin{figure}
\includegraphics[width=9.0cm]{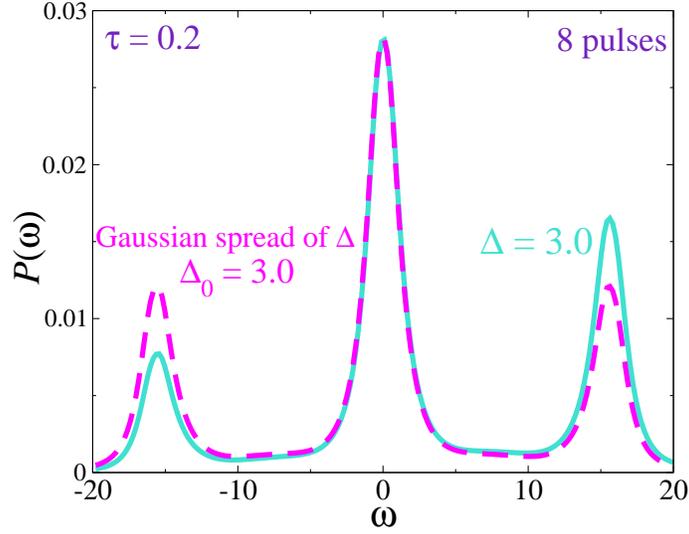} 
\caption{(Color online) Comparison of the pulse-controlled spectra between the single emitter with $\Delta=3$ (solid cyan line) and the inhomogeneously broadened ensemble of emitters with the Gaussian distribution of the detunings $\Delta$ having the standard deviation $\Delta_0=3$ (dashed magenta line). As expected, the pulse control suppresses the broadening and makes the satellite peaks symmetric.} 
\label{SIfigMEqtn_GaussDelta}
\end{figure}

We also present the pulse-controlled spectrum, calculated with the master equation approach, for an inhomogeneously broadened {\it ensemble\/} of emitters in Fig.~\ref{SIfigMEqtn_GaussDelta} (dashed magenta line). We assumed that each individual emitter has a random (quasi)static value of $\Delta$, distributed according to Gaussian law $P(\Delta)=(1/\sqrt{2\pi\Delta_0^2})\exp{[-\Delta^2/(2\Delta_0^2)]}$, with the deviation $\Delta_0=3$, and all other parameters are the same as in Fig.~\ref{SIfigAllResCompare}. For comparison, we also present the pulse-controlled spectrum for a single emitter with $\Delta=3$ (solid cyan line), i.e.\ the same result as shown in Fig.~\ref{SIfigAllResCompare} with cyan line. As expected, the pulse control efficiently suppresses the broadening: the central peak in both spectra is the same, and the only difference is in the heights of the sattelites. This is exactly the expectation: the symmetric distribution of $\Delta$ makes the satellite peaks symmetric (crudely speaking, averaging the left and the right satellites from Fig.~\ref{SIfigAllResCompare}).

\subsection{Two-photon interference (TPI) experiments}

To assess the photon indistinguishability, we explicitly calculate the coincidence rate in the TPI experiments. We assume that the photons from two emitters, with the detunings $\Delta_1$ and $\Delta_2$, arrive to the input modes (described with the photon creation operators $a^\dag_1$ and $a^\dag_2$) of a 50:50 beamsplitter, and two detectors count the photons at the output modes described by the creation operators $a^\dag_3$ and $a^\dag_4$:
\begin{equation}
a^\dag_3=(1/\sqrt{2})[a^\dag_1+a^\dag_2],\quad a^\dag_4=(1/\sqrt{2})[a^\dag_2-a^\dag_1].
\end{equation}
The probability that the two detectors click at times $t$ and $t+\theta$, respectively, is (see e.g.\ Ref.~\onlinecite{KirazEtal})
\begin{equation}
P_{34}(t,\theta)=\langle a^\dag_3(t) a^\dag_4(t+\theta) a_4(t+\theta) a_3(t)\rangle,
\end{equation}
and can be expressed via the correlation functions of the emitters' operators as:
\begin{equation}
P_{34}\propto (1/4)[G_1^{(2)}(t,\theta) + G_2^{(2)}(t,\theta)] + G_{34}(t,\theta),
\end{equation}
where $G_1^{(2)}(t,\theta)=\langle\sigma_1^+(t)\sigma_1^+(t+\theta)\sigma_1^-(t+\theta)\sigma_1^-(t)\rangle$ (and similarly for $G_2^{(2)}$) are the single-emitter terms. These terms are often either small or are omitted, see e.g.\ Ref.~\cite{KirazEtal}, and do not measure indistinguishability of the photons from different emitters.

The part which determines the photon indistinguishability is the term describing the two-photon interference between different emitters:
\begin{equation}
G_{34}(t,\theta)=(1/4)\left[\phi_1(t+\theta,0) \phi_2(t,0) + \phi_1(t,0) \phi_2(t+\theta,0) - \phi_1(t,\theta)^* \phi_2(t,\theta) - \phi_1(t,\theta) \phi_2(t,\theta)^*\right],
\label{tpiG34}
\end{equation}
which is determined by the correlation functions (see Eq.~\ref{cf} above) of the emitters. 
The normalized TPI term is
\begin{equation}
g_{34}(t,\theta)=G_{34}(t,\theta)/N_{34}(t,\theta),
\end{equation}
where
\begin{equation}
N_{34}(t,\theta)= (1/4)\left[\phi_1(t,0) + \phi_2(t,0)\right]\  \left[\phi_1(t+\theta,0) + \phi_2(t+\theta,0)\right].
\end{equation}

Without pulses, $\phi_1(t,\theta)=\exp{[-\Gamma t+(i\Delta_1-\Gamma/2)\theta]}$, and, similarly, for $\phi_2$; we assume here for simplicity that both emitters have the same natural linewidth $\Gamma$. In this case $G_{34}=[\exp{(-2\Gamma t-\Gamma\theta)}] [1-\cos{(\theta\Delta_{21})}]/2$, where 
\begin{equation}
\Delta_{21}=\Delta_2-\Delta_1.
\end{equation} 
If the emitters' frequencies undergo random spectral diffusion, the oscillating term averages out to zero, and we have TPI between two independent sources. The normalized count $g_{34}=\sin^2{(\Delta_{21}\theta/2)}$ under this condition also averages out to 1/2.

To analyze the situation with pulses, for simplicity, let us focus on the long-time limit, where stationary regime is achieved: in this case, for $\Gamma\tau\ll 1$, we have $\rho_e(t)\approx 1/2$, and the TPI term is
\begin{equation}
G_{34}(t,\theta) = (1/8)\left[1 - {\rm Re} f_1(t,\theta) f_2(t,\theta)^*\right],
\end{equation}
and the normalization factor is just $N_{34}=1/4$. The behavior of the function $f(t,\theta)$ has been analyzed above (Eq.~\ref{ffunc} and the discussion following it). When the times $t$ and $t+\theta$ are separated by even number of pulses, we find that
\begin{eqnarray}
G_{34}(t,\theta)&=& (1/8)\left(1-\cos{[\Delta_{21}(\tau_1+\tau_2-\tau)]}\right),\\
g_{34}(t,\theta)&=& (1/2)\left(1-\cos{\left[\Delta_{21}(\tau_1+\tau_2-\tau)\right]}\right),
\end{eqnarray}
and the cosine term is always close to 1 because $\Delta_{21}(\tau_1+\tau_2-\tau)\ll 1$ for sufficiently small inter-pulse delay.
Thus, in this case we have TPI between two almost perfectly coherent emitters, as if the spectral diffusion was absent, and the TPI term remains almost zero. When $t$ and $t+\theta$ are separated by odd number of pulses, the values of $G_{34}$ and $g_{34}$ have the values 1/8 and 1/2, respectively, as in the case of independent dephased sources. This is exactly our prediction based on the spectral shape: we see that for TPI also about half of the photons become indistinguishable under the action of the pulses. Thus, we explicitly see that the pulses do significantly improve the indistinguishability of the emitted photons.

Experimentally, one can screen out the events where the two clicks are separated by odd number of pulses, or settle for average $g_{34}=1/4$, where half of the photons correspond to the suppressed spectral diffusion. 

For completeness, we also calculated the single-emitter 2nd-order intensity correlators $G_1^{(2)}$ and $G_2^{(2)}$. They are independent of the emitter's frequency, and are identical for both emitters. When $t$ and $t+\theta$ are separated by the even number of pulses, i.e. when $\theta=2k\tau+\theta_1$ with $\theta_1\in[0,\tau)$, we have $G^{(2)}\approx (1/4)(1-e^{-\Gamma\theta})$. When the two instants are separated by the odd number of pulses, $G^{(2)}\approx (1/4)(1+e^{-\Gamma\theta})$. Again, one can screen the events corresponding to the odd nubmer of pulses, and have negligible $G^{(2)}$ for $\theta\ll t_0=1/\Gamma$; this case is similar to the case of the resonance fluorescence with weak driving \cite{Loudon,MattAtaturePRLResFl}.

If $t$ is not a controlled parameter, and only the delay $\theta$ between the two clicks is recorded, then the relevant quantities should be averaged over time. In the limit of small inter-pulse delay, in the leading order in $\tau$, the answers are easily formulated. When the two clicks are separated by an even number of pulses, i.e.\ when $\theta=2j\tau+\theta_1$ with $\theta_1\in[0,\tau)$, we get
\begin{equation}
{\bar g}_{34}(\theta)= \frac{1}{2}\ \frac{\theta_1}{\tau},
\end{equation}
where the overbar means averaging over time $t$. for the odd number of pulses, when $\theta=(2j+1)\tau+\theta_1$ with $\theta_1\in[0,\tau)$, we get
\begin{equation}
{\bar g}_{34}(\theta) = \frac{1}{2}\ \left[1-\frac{\theta_1}{\tau}\right].
\end{equation}
Again, the significant improvement over the no-pulse case is clearly seen. One can screen the events corresponding to the values of $\theta$ which are close to odd integers of $\tau$, thus decreasing ${\bar G}_{34}$ and ${\bar g}_{34}$, or settle with the average-case scenario, with ${\bar g}_{34}=1/4$.

\section{Brief discussion of other control protocols}

In this work we focus on one control sequence, which consists of the waiting time $\tau$ and subsequent $\pi$-pulse along the $x$-axis; this sequence is often known as a periodic dynamical decoupling (PDD) protocol, denoted often as $[\tau-\pi-\tau-\pi]^{N_p/2}$ where $N_p$ is the total number of pulses. 

In the area of dynamical decoupling (DD), a number of other protocols has been developed. Detailed analysis of various DD protocols is beyond the scope of this  work. The goal of the present study is to show that even such a fast environment as photons can be controlled with realistic pulses. Indeed, the photon reservoir has practically no memory, responding almost instantly (on a timescale of order of the optical oscillation period $2\pi/\omega$, i.e.\ femtoseconds) to the changes in the emitter's state. The fact that the nanosecond-scale pulses can efficiently control such fast a system is of interest by itself; this result is also of utmost importance for developing quantum networks with solid-state emitters. Considering one exemplary sequence which achieves the desired control is sufficient for this purpose.

However, it might be of interest to briefly discuss other sequences and the relation of our approach to DD in some detail. 

The standard DD considers a system with finite memory time, subjected to a sequence of pulses which are applied on a timescale smaller or of the order of the system's memory time; for simplicity let us focus on ideal (instantaneous 180$^\circ$) pulses. Due to finite memory time, the system's evolution after a pulse is related to its evolution before a pulse. If the pulse sequence is chosen appropriately, the overall system's evolution (unitary or non-unitary) after the sequence of pulses is modified, as if the system evolved under the action of some Hamiltonian/Liouvillian which differs from the original Hamiltonian/Liouvillian in the absence of control.

At first sight, our problem seems similar to the standard DD. We apply pulses to the emitter, in order to change its evolution in such a manner that the emitter's motion would not be affected by the detuning term $(\Delta/2)\sigma_z$. Since the emitter's response time is finite, being governed by the parameters $\Delta$ and $\Gamma$, we should apply the pulses on a timescale $\tau$ much less than $1/\Delta$ and $1/\Gamma$ to achieve that.

However, the actual problem we study here, control of the fast photon reservoir, is principally different in several aspects. First, note that our goal is not to control the emitter itself: although we apply pulses to the emitter, our actual goal is to control the emitted photons. As mentioned above, the photons are qualitatively different from the standard DD systems: the photonic reservoir has practically no memory, and reacts practically instantly to the changes in the emitter's state. Therefore, the emission process, i.e. the transfer of the emitter's state into the state of the photon bath, happens during all times, and the emission at the time $t$ is controlled by the state of the emitter at the time $t$ (up to the delay $r_s/c$); in a formal way this is seen from Eqs.~\ref{rhoevol} and \ref{aVSsigma}. What the pulses affect is an integral over different elementary emission steps, happening at different times, as Eq.~\ref{longtimeP} demonstrates. Similarly, the TPI coincidence count rates, Eq.~\ref{tpiG34} and below, are determined by different elementary emission steps, happening at different times. Thus, in our case, the pulses affect the correlations between different emission steps happenning at different times, and all (or at least a large fraction) of such steps is important.

The second principal difference is closely related to the first one. In standard DD, the pulses are applied to make the system move as if its evolution were governed by some desired Liouvillian ${\cal L}_0$ instead of its original (control-free) Liouvillian $\cal L$. Obviously, this cannot happen at all times: even if at time $t_1$ we achieved ideal decoupling, such that the system's evolution (super)operator in the presense of control ${\cal E}_c(t_1)$ exactly coincides with the desired one ${\cal E}_0(t_1)$, this ideal equality will be destroyed the next moment. Indeed, until the next pulse arrives, the system's evolution will proceed in a control-free manner, i.e.\ for the moment of time $t_2$ between $t_1$ and the arrival of the next pulse, ${\cal E}_c(t_2) = {\cal E}(t_2,t_1) {\cal E}_c(t_1)$, where ${\cal E}(t_2,t_1)$ describes the control-free evolution of the system between times $t_1$ and $t_2>t_1$. 

In the standard DD settings, this issue is of little relevance, as  the condition ${\cal E}_c(t_1)\approx {\cal E}_0(t_1)$ is satisfied only for some fixed moments of time $t_1$, when the system's evolution is ``refocused" by pulses. This freedom is often used to re-arrange the timings of the pulses, to make the controlled evolution ${\cal E}_c(t_1)$ as close as possible to the desired ${\cal E}_0(t_1)$, and often the only relevant moment of time is the end of the sequence, i.e.\ $t_1=T$. The pulses are supposed to be applied sufficiently frequently, with the typical inter-pulse distance $T/N_p$ (where $N_p$ is the total number of pulses in the sequence) small enough to ensure $||{\cal L}(T)-{\cal L}_0(T)|| T \ll 1$ \cite{LZK}, and different sequences can be classified based on the suitable expansion of the difference $||{\cal E}_c(T)-{\cal E}_0(T)||$ in terms of the small parameter $T/N_p$. For instance, the PDD sequence in our work is of the first order, while the often-used symmetrized Carr-Purcell (CP) sequence $[(\tau/2)-\pi-\tau-\pi-(\tau/2)]^{N_p/2}$ is of the second order, and the so-called Uhrig's DD sequence (UDD), where the pulses are timed according to zeros of a sine function, has the order $N_p$.

In our case, such a notion of the degree of decoupling is of little relevance: the elementary emission steps happen at all times, and the resulting emission spectra and the relevant correlation functions are not determined by the emitter's state only near some specific refocusing time instants. As we mentioned above, for a pulse-controlled system, there is always a control-free evolution during some times, at least during the interval between the refocusing time and the arrival of the next pulse. Thus, there is always an additional undesired contribution to the emitter's correlation function $\phi(t,\theta)$ of the order of $T/N_p$, and the corresponding contributions to the emission spectrum and the TPI correlators. 

In fact, our situation is even more different from the standard DD: as one can see in the previous Section, for any pulse sequence the correlation function of the emitter $\phi(t,\theta)$ is zero if the number of pulses between the instants $t$ and $t+\theta$ is odd. This leads to appearance of the satellite peaks in the emission spectrum and to an increase in the coincidence count rate during odd intervals. Thus, the evolution of the photon reservoir close to the desired one (such as directing all emission into the central peak near $\omega=0$) is not achievable by any pulse sequence. The 50\% of indistinguishable photons, achieved with PDD, is close to the realistically achievable maximum (qualitatively speaking, half of the time should be spent during odd pulse intervals), but corresponding detailed analysis is beyond the scope of this paper, and will be presented elsewhere.

\begin{figure}
\includegraphics[width=9.0cm]{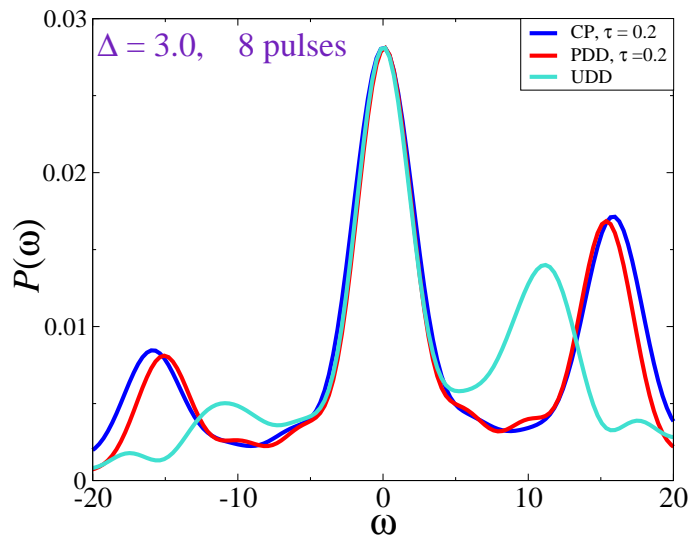} 
\caption{(Color online) The pulse-controlled spectra of the emitter with $\Delta=3.0$ subjected to the PDD, CP, and UDD sequences with total time $T=1.6$ and total number of pulses $N_p=8$. As expected, all sequences produce very similar central peak; also, CP and PDD give very close spectral profiles, while the satellites for UDD are quite different due to different time modulation of the emitter's correlation function. All curves were obtained by numerically integrating Eq.~\ref{rhoevol} and calculating the integral Eq.~\ref{pintegr}.} 
\label{SIfigDDprotocols}
\end{figure}

Thus, the standard comparison of the DD pulse sequences, based on the order of decoupling or on some other measure which assumes that $||{\cal E}_c(T)-{\cal E}_0(T)||$ can be very close to zero, is not applicable to our problem: these two quantities are never close even for very small inter-pulse delays. However, different sequences provide different time profiles of modulation of the correlation function $\phi(t,\theta)$, and therefore lead to different modifications of the emission spectrum. This can be exploited if one wants to achieve different spectral profiles, and would be of great interest for applications; it would constitute an excellent direction for future research. 
Another interesting possibility is to investigate continuous-wave decoupling, and, more generally, control of the emitter with arbitrary time-dependent control fields. 

A good illustration of this thesis is the comparison of three sequences: PDD, CP, and UDD, shown in Fig.~\ref{SIfigDDprotocols}. All curves were obtained by numerically integrating Eq.~\ref{rhoevol} and calculating the integral Eq.~\ref{pintegr}.
The master equation analysis for the CP sequence shows that the emission spectrum and the TPI coincidence count functions are almost the same as for the PDD sequence, differing only by small terms of order of $\tau\Delta$ and $\Gamma\tau$, and the simulations based on master equation explicitly show that. I.e., although the CP sequence formally has higher order, it leads to practically the same results as PDD. Similarly, UDD, despite its high-order decoupling, does not direct all photons to the central emission peak. However, different time modulation of the emitter's correlation function leads to a different spectral profile of emission.

\begin{figure}
\includegraphics[width=9.0cm]{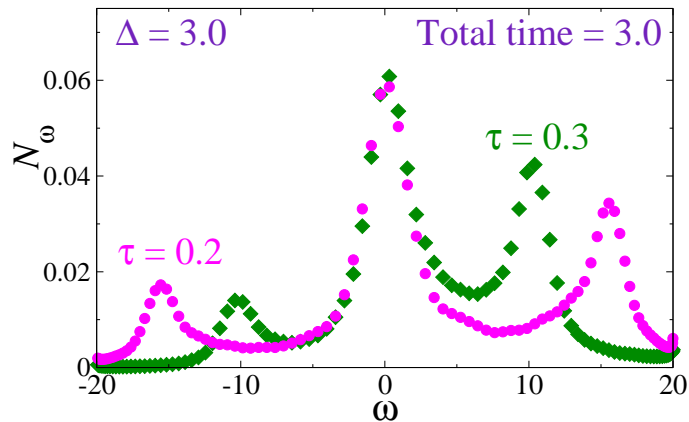} 
\caption{(Color online) The pulse-controlled spectra of the emitter with $\Delta=3.0$ subjected to the sequences with $\tau=0.2$ (magenta dots) and $\tau=0.3$ (green diamonds). For both spectra the total time is $T=3.0$, and the amplitude and the shape of the central peak is the same for both sequences, although the satellite peaks are positioned differently.} 
\label{SIfigTotalTime3}
\end{figure}

Also note that the amplitude of the central emission peak, as predicted by the master equation analysis above, has the total time $T$ as a pre-factor. This prediction  is corroborated by the tDMRG simulations shown in Fig.~\ref{SIfigTotalTime3}, where the simulation results are shown for the emitter with $\Delta=3.0$ subjected to the control sequences with $\tau=0.2$ (magenta dots) and $\tau=0.3$ (green diamonds). The spectra for both control sequences are presented for the same total time $T=3.0$, and the amplitude and the shape of the central peak is the same for both sequences, although the satellite peaks are positioned differently.